\documentclass[a4paper,11pt]{article}
\pdfoutput=1 

\usepackage{jheppub} 
                     
\usepackage{amsmath,amssymb,amsthm,amscd,graphicx}
\usepackage{psfrag}
\input epsf.sty

\usepackage{color}

\theoremstyle{definition}

\newcommand{\CA}{{\cal A}}
\newcommand{\CB}{{\cal B}}

\newcommand{\acycle}{{\cal A}}
\newcommand{\bcycle}{{\cal B}}
\newcommand{\fad}{\operatorname{\Phi}_{\mathsf{b}}}

\newcommand{\mypsi}[2]{\operatorname{\Psi}_{#1,#2}}

\newcommand{\im}{{\mathsf{i}}}

\def\IZ{{\mathbb Z}}
\def\IR{{\mathbb R}}
\def\IC{{\mathbb C}}
\def\IP{{\mathbb P}}

\newcommand{\re}{{\rm e}}
\newcommand{\ri}{{\rm i}}
\newcommand{\rd}{{\rm d}}

\newcommand{\Tr}{\mathop{\rm Tr}\nolimits}

\newcommand{\lt}{\left(}
\newcommand{\rt}{\right)}
\newcommand{\be}{\begin{equation}}
\newcommand{\ee}{\end{equation}}
\newcommand{\ba}{\begin{aligned}}
\newcommand{\ea}{\end{aligned}}
\newcommand{\ben}{\begin{eqnarray}\displaystyle}
\newcommand{\een}{\end{eqnarray}}

\newdimen\tableauside\tableauside=1.0ex
\newdimen\tableaurule\tableaurule=0.4pt
\newdimen\tableaustep
\def\phantomhrule#1{\hbox{\vbox to0pt{\hrule height\tableaurule width#1\vss}}}
\def\phantomvrule#1{\vbox{\hbox to0pt{\vrule width\tableaurule height#1\hss}}}
\def\sqr{\vbox{%
  \phantomhrule\tableaustep
  \hbox{\phantomvrule\tableaustep\kern\tableaustep\phantomvrule\tableaustep}%
  \hbox{\vbox{\phantomhrule\tableauside}\kern-\tableaurule}}}
\def\squares#1{\hbox{\count0=#1\noindent\loop\sqr
  \advance\count0 by-1 \ifnum\count0>0\repeat}}
\def\tableau#1{\vcenter{\offinterlineskip
  \tableaustep=\tableauside\advance\tableaustep by-\tableaurule
  \kern\normallineskip\hbox
    {\kern\normallineskip\vbox
      {\gettableau#1 0 }%
     \kern\normallineskip\kern\tableaurule}%
  \kern\normallineskip\kern\tableaurule}}
\def\gettableau#1{\ifnum#1=0\let\next=\null\else
\squares{#1}\let\next=\gettableau\fi\next}

\newcommand{\figref}[1]{Fig.~\protect\ref{#1}}

\title{\boldmath  Spectral determinants and quantum theta functions }
\author{ Alba Grassi
}

\affiliation{International Center for Theoretical Physics,\\
 ICTP-UNESCO, Strada Costiera 11, Trieste 34151,Italy \\

}

\emailAdd{agrassi@ictp.it}

\abstract{
It has been recently conjectured that the spectral determinants of operators associated to mirror curves 
can be expressed in terms of a generalization of theta functions, called quantum theta functions. 
In this paper we study the symplectic properties of these spectral determinants by expanding them around the point $\hbar=2\pi$, where the quantum theta functions become conventional theta functions. We find that they are modular invariant, order by order, and we give explicit expressions for the very first terms of the expansion.
 Our derivation requires a detailed understanding of the modular properties of topological string free energies in the Nekrasov--Shatashvili limit. We derive these properties in a diagrammatic form. Finally, we use our results  to provide a new test of the duality between topological strings and spectral theory.
}

\begin{document}
\maketitle
\flushbottom

\section{Introduction}

Topological string theory was introduced in the '80s as a simplified model of string theory which captures  information on the background geometry \cite{ts1}. Since then, this theory has led to many new results in mathematical physics. 

Originally this model was defined only perturbatively by a formal asymptotic expansion. When the background geometry is a toric Calabi-Yau (CY), one can compute the coefficients of this series at any order in perturbation theory~\cite{akmv2,ko,kz,bcov}. Many aspects of the large order behavior of this formal expansion have been studied in a series of papers starting with the seminal paper~\cite{mmopen}  and leading to several insights on a possible non--perturbative definition of the theory \cite{gikm, msw1,ps, dmpnp, mmnp, asv, cesv1,mpp, cesv2, gmz}. 
Recently, by using the so--called Fermi gas formalism, a non--perturbative completion  of topological strings on toric CY has been proposed. This formalism was first introduced in the context of ABJM theory \cite{mp} and further extended to topological string theory  in \cite{ghm}, by using the insights from \cite{hmmo,km}.

In  the non--perturbative formulation of \cite{ghm} 
one  associates a trace class, positive definite operator, $  \hat \rho_X $, to any toric Calabi-Yau 3-fold $X$ whose mirror geometry has genus one (the higher genus generalization was later worked out in \cite{cgm2}) . Physically, this operator corresponds to the  density matrix of an ideal Fermi  gas with a positive and discrete spectrum.  Such a spectrum  can be elegantly organized into the so--called spectral (or Fredholm)  determinant
\be \Xi({ \kappa}, \hbar)=\det \left( 1+\kappa \hat \rho_X\right).\ee
In  \cite{ghm} it has been conjectured an explicit expression for this spectral determinant in terms of (un)refined Gopakumar-Vafa  invariants of X.  At the orbifold point ($\kappa=0$), the spectral determinant, denoted by $ \Xi_X(\kappa, \hbar)$, has the following expression 
  \be \label{skappa} \Xi({ \kappa}, \hbar)=1+\sum_{N\geq 1}\kappa^N Z(N, \hbar),\ee
 where  $Z(N, \hbar)$ can be identified with the partition function of a one dimensional ideal Fermi gas with N particles. 
 The proposal of \cite{ghm}  is that $Z(N, \hbar)$
   provides a non-perturbative definition for the topological string partition function where the string coupling $g_s$ is identified with the inverse of the Planck constant $\hbar$. This gives a duality between topological string theory and quantum mechanics similarly  to the AdS/CFT correspondence \cite{adscft}  where on the AdS side we have topological string theory on toric CYs  while the role of the CFT is played by a one dimensional Fermi gas.

   In this formalism the perturbative expansion of standard (or unrefined) topological strings emerges when we study the t' Hooft limit of the gas, namely
    \be\label{thl} N, \hbar \to \infty, \qquad {N\over \hbar} \quad \text{fixed}.\ee
  However,  there are important non--perturbative corrections to the t' Hooft  limit \eqref{thl} which can be detected by looking at the thermodynamic limit of the gas
         \be N \to \infty, \qquad { \hbar} \quad \text{fixed}.\ee
         These non-perturbative corrections are determined by  the so--called  Nekrasov--Shatashvili (NS) limit of the refined topological string \cite{hmmo, ghm}.
 By combining  the results of \cite{ghm,kama} with the Cauchy identity \cite{mp}, one can  write $Z(N, \hbar)$ as a matrix model whose 't Hooft expansion 
 reproduces the perturbative expansion of topological strings in the conifold frame \cite{mz,kmz}.

The explicit expression for the spectral determinant  in terms of enumerative invariants  proposed in \cite{ghm,cgm2} is based on an expansion at the large radius point of the moduli space where 
\be \kappa \gg 0.\ee
However, to fully understand $\Xi(\kappa, \hbar)$ at a generic point of the moduli space one would need a Gopakumar-Vafa-like resummation \cite{gv} away from the large radius point. Although,  there are evidences and concrete examples to believe that such a resummation should exist \cite{ghmabjm, oz}, at present it is not known.  %
Nevertheless, there are some special cases, called the "maximally supersymmetric"  cases, where  it is possible to  write down a closed form expression for the spectral determinant in terms of hypergeometric and conventional theta functions \cite{cgm,ghm}. In these cases we have a good control of the spectral determinant over the full moduli space. Such a simplification occurs for instance when $\hbar=2 \pi$. 
Therefore in our formalism the simplest regime is a quantum regime where $\hbar=2\pi$ and not the classical one.

The strategy in this paper is to study $\Xi(\kappa,\hbar)$ at a generic point of the moduli space (parametrized by $\kappa$) by performing an expansion around the special value $\hbar=2\pi$
\be\label{exp}  \Xi ( {\boldsymbol \kappa},\hbar)=\sum_{n\geq 0}{1\over n!}{\rd^n \Xi ( {\boldsymbol \kappa}, 2\pi) \over \rd^n \hbar}(\hbar-2\pi)^n. \ee
At each order in the expansion one can write down a closed form expression for the spectral determinant in term of unrefined and NS free energies where the  $n^{\rm th}$ order  is completely determined
by the free energies $F_g$ and $F_h^{\rm NS}$ with $g, h\leq n$. In this paper  we will work out in details the very first few terms in this expansion \eqref{exp}.

It  is well known \cite{we,abk} that different points in the moduli space correspond to different  choices of  symplectic frames.
The transformations properties of the unrefined free energies under such a change of frame have been worked out in detail in \cite{we,abk} where it was shown that the unrefined free energies transform as a wave function while moving around  the moduli space.
In this paper we will see that a similar behavior holds also for the NS free energies. By using these considerations we argue that,  order by order in the $\hbar-2\pi$ expansion \eqref{exp}, the spectral determinant is  invariant under a change of symplectic frame.

This paper is organized as follows. In  sections 2 and 3 we  review some of the results concerning the special geometry of toric Calabi-Yaus and the conjecture of \cite{ghm}. In section 4 we  organize the modular properties of the Nekrasov-Shatashvili free energies by using diagrammatic rules and  in section 5 we study the expansion of the spectral determinant around $\hbar=2\pi$. In section 6 we use these results to give further evidence for the topological string/quantum mechanics duality proposed in \cite{ghm}.  %
 \section{Special Geometry}
We consider a  toric Calabi Yau 3-fold X whose mirror geometry is encoded in a Riemann surface of genus $g_\Sigma$:
\be \label{sf}W_X(\re^{x}, \re^{p}, {\boldsymbol \kappa})=0, \quad x,p \in \IC,\ee 
where \be {\boldsymbol \kappa}=\left\{\kappa_1,\cdots ,\kappa_{g_\Sigma} \right \},\ee
are the complex deformation parameters describing the geometry \eqref{sf}, see \cite{hkrs,hkp} for more details.  

Let  $ \mathcal{A}_I,\mathcal{B}_I$ be a symplectic basis of cycle for the above surface
\be\label{basc}
{\acycle}_I \cap {\bcycle}_J \propto  \delta_{IJ}, 
\qquad 
{\acycle}_I \cap{\acycle}_J=0, 
\qquad
 {\bcycle}_I \cap {\bcycle}_J=0, \qquad  I,J = 1, \cdots,  g_\Sigma.
\ee
The classical periods for the geometry \eqref{sf} are given by
\be \label{clp}\Pi_{A_I}( {\boldsymbol \kappa})=\oint_{{\acycle}_I }p(x){\rd x}, \quad  \Pi_{B_I}( {\boldsymbol \kappa})=\oint_{{\bcycle}_I }p(x){\rd x},\ee
where $p(x)$ is determined by \eqref{sf}.
For sake of simplicity we omit  the dependence on ${\boldsymbol \kappa}$ in $p(x)$.
The classical mirror map  $\epsilon_I$ and  the genus zero free energy $F_0$  of topological string on X, can be computed from these periods as \cite{ckyz}
\be \label{cl} \epsilon_I( {\boldsymbol \kappa})=- \Pi_{A_I}( {\boldsymbol \kappa}), \qquad \partial_{\epsilon_I}F_0( {\boldsymbol \kappa})=-\Pi_{B_I}( {\boldsymbol \kappa}).\ee
Sometimes we refer to $\epsilon_I$ as classical K\"ahler parameter.
When performing a modular transformation of the cycles
\be\label{cytra}\left( \begin{array}{c}
{\CB} \\ 
{\CA}
\end{array} \right) \rightarrow \left( \begin{array}{cc}
A & B \\ 
C & D
\end{array} \right) \left( \begin{array}{c}
{\CB }\\ 
{\CA}
\end{array} \right),  \quad \begin{pmatrix}A& B \\
               C&D \end{pmatrix}  \in {\rm Sp}(2 g, \IZ) \ee 
the classical periods transform according to \cite{abk}
\be \label{cltra}
\left( \begin{array}{c}{\Pi_{B}} \\ 
{\Pi_{A}}
\end{array} \right) \rightarrow \left( \begin{array}{cc}
A & B \\ 
C & D
\end{array} \right) \left( \begin{array}{c}
{ \Pi_{B}}\\ 
{\Pi_{A}}
\end{array} \right).\ee
 This is what it is called a change of symplectic frame. In the following we will use the term symplectic  (or modular) transformation to refer to  such a change of  frame.
 
In \cite{acdkv}, based on earlier works  \cite{adkmv,mirmor,mirmor2,mt,Bonelli:2011na}, it was argued that one can  quantize the curve \eqref{sf} by promoting $x$ and $p$ to operators fulfilling the standard commutation relation
\be \left[ \hat x, \hat p \right]=\ri \hbar .\ee
The differential $ p(x)\rd x$
is then promoted to a quantum differential \be p(x, \hbar)\rd x \ee
fulfilling
\be W(\re^{\hat x}, \re^{\hat  p}, {\boldsymbol \kappa})\left(\exp \left[{{1\over \hbar}\int^x p(y, \hbar)\rd y}\right]\right)=0. \ee
The corresponding quantum periods are defined by
\be  \Pi_{A_I}(\hbar, {\boldsymbol \kappa})=\oint_{{\acycle}_I }p(x,\hbar){\rd x}, \quad  \Pi_{B_I}(\hbar,  {\boldsymbol \kappa})=\oint_{{\bcycle}_I }p(x,\hbar){\rd x}.\ee
These periods are related to the so-called  Nekrasov--Shatashvili (NS) free energy \eqref{Fns} through
\be \label{QP} \epsilon_I(\hbar,  {\boldsymbol \kappa})=-\Pi_{A_I}(\hbar, {\boldsymbol \kappa}), \qquad  \hbar\partial_{ \epsilon_I}F^{\rm NS}( {\hbar,\boldsymbol \kappa})=-\Pi_{B_I}(\hbar, {\boldsymbol \kappa}).\ee
In \cite{hkrs}, based on earlier work \cite{huangNS,mirmor}, it was argued that
\be \label{qps} \ba 
\Pi_{A_I}(\hbar, {\boldsymbol \kappa})=& \sum_{n\geq 0}\Pi_{A_I}^{(n)}( {\boldsymbol \kappa})\hbar^{2n}, \qquad \Pi_{B_I}(\hbar, {\boldsymbol \kappa})=& \sum_{n\geq 0} \Pi_{B_I}^{(n)}( {\boldsymbol \kappa})\hbar^{2n}\ea \ee
with
\be \label{huang}\ba  \Pi_{A_I}^{(n)}( {\boldsymbol \kappa})&= \mathsf{D}_1^{(n)}( {\boldsymbol \kappa})\Pi_{A_I}^{(0)}+ \mathsf{D}_2^{(n)}( {\boldsymbol \kappa})\Pi_{A_I}^{(0)}, \\
\Pi_{B_I}^{(n)}( {\boldsymbol \kappa})&= \mathsf{D}_1^{(n)}( {\boldsymbol \kappa})\Pi_{B_I}^{(0)}+ \mathsf{D}_2^{(n)}( {\boldsymbol \kappa})\Pi_{B_I}^{(0)}. \ea\ee
where \be \Pi_{A,B}^{(0)}=\Pi_{A,B}( {\boldsymbol \kappa})\ee
denote the classical periods \eqref{clp} and   $ \mathsf{D}_i^{(n)}( {\boldsymbol \kappa})$ are differential operators of order $i$. We  give a concrete example in appendix \ref{soid}.
It follows that the quantum periods transform according to
\be \label{pt}\left(\begin{matrix} \hbar \partial_{{\boldsymbol \epsilon} } F^{\rm NS} \\  {\boldsymbol \epsilon} \end{matrix}\right) \to  \left(\begin{matrix} A & B \\ C & D \end{matrix}\right)  \left(\begin{matrix} \hbar \partial_{{\boldsymbol \epsilon} } F^{\rm NS} \\ {\boldsymbol \epsilon}  \end{matrix}\right).\ee
To make contact with the results of \cite{ghm,cgm2} it is useful to introduce
\be \ba \label{khalerdef} {t}_I ( {\boldsymbol \kappa},\hbar)=& -{1\over (2\pi \ri)^{1/2}}\mathcal{C}_{IJ} \epsilon_J ( {\boldsymbol \kappa},\hbar) , \quad I,J=1, \cdots g_\Sigma .\ea \ee
The matrix $\mathcal{C}_{IJ}$  is a $g_\Sigma \times g_\Sigma$ matrix which can be read off from the toric data of $X$ as explained in \cite{cgm2,kpsw}.
We also use
\be\ba
 {T}_I ( {\boldsymbol \kappa},\hbar)=& {1\over (2\pi \ri)^{1/2}}\epsilon_I ( {\boldsymbol \kappa},\hbar) , \quad I,J=1, \cdots g_\Sigma. \ea \ee
It follows that
\be \label{pt2}\left(\begin{matrix} \hbar \partial_{{\bf t} } F^{\rm NS} \\ 2\pi \ri {\bf t} \end{matrix}\right) \to  \left(\begin{matrix} A & B \\ C & D \end{matrix}\right)  \left(\begin{matrix} \hbar \partial_{{\bf t} } F^{\rm NS} \\ 2\pi \ri {\bf t}  \end{matrix}\right),\ee
\be \label{pt2b}\left(\begin{matrix} \hbar \partial_{{\bf T} } F^{\rm NS} \\ 2\pi \ri {\bf T} \end{matrix}\right) \to  \left(\begin{matrix} A & B \\ C & D \end{matrix}\right)  \left(\begin{matrix} \hbar \partial_{{\bf T }} F^{\rm NS} \\ 2\pi \ri {\bf T}  \end{matrix}\right).\ee
Depending on the kind of computation it is more convenient to use ${\bf T }, {\bf t }$ or $\boldsymbol \epsilon$.
\section{Spectral determinant and topological strings} 

In this section we will review some  aspects of the work \cite{ghm} and its generalization to mirror curves of higher genus  \cite{cgm2}.
A pedagogical review of this conjecture and its implications has been presented in \cite{mmnew}.  

The starting point is the quantization of the mirror curve \eqref{sf} which leads to $g_\Sigma$
trace class operators (see \cite{kama} for a rigorous mathematical proof) \be \label{rhoop}\hat \rho_i( \tilde{\boldsymbol \kappa}_i), \quad i=1, \cdots, g_\Sigma ,\ee
where
\be\label{kappa}\tilde{\boldsymbol \kappa}_i=\left\{\kappa_1, \cdots, \kappa_{i-1},\kappa_{i+1},\cdots,\kappa_{g_\Sigma} \right\}. \ee
The operators \eqref{rhoop} have a discrete spectrum which can be elegantly organized into a generalized spectral determinant 
\be \label{gsd} \Xi_X (  {\boldsymbol \kappa}, \hbar)= {\rm det}\left(1+ \kappa_1 \hat \rho_1\right) {\rm det}\left(1+ \kappa_2 \hat \rho_2\Big|_{\kappa_1=0} \right) 
 \cdots {\rm det}\left(1+ \kappa_{g_\Sigma} \hat \rho_{g_\Sigma} \Big|_{\kappa_1=\cdots = \kappa_{g_\Sigma-1}=0} \right). \ee
 More precisely, the spectrum of the operators $\hat \rho_i( \tilde{\boldsymbol \kappa}_i)$ is determined by vanishing locus of  $\Xi_X ( {\boldsymbol \kappa}, \hbar)$ as explained in \cite{cgm2}.
 In the case of a genus one mirror curve the generalized spectral determinant \eqref{gsd} becomes the standard Fredholm determinant 
 \be \label{sd} \Xi_X(\kappa, \hbar)=\det(1+\kappa \hat \rho_X)=\prod_{n}\left(1+\kappa \re^{-E_n}\right),\ee
where  $\re^{-E_n}$ are the eigenvalues of  $ \hat \rho_X$.

As an example we consider the anti canonical bundle over $\mathbb{CP}^2$. 
For this geometry the mirror curve \eqref{sf} reads
\be W(x,p,\kappa)=\re^x +\re^{p}+\re^{-x-p}+\kappa, \ee
and the corresponding quantum operator is
\be \label{thop2} \hat \rho_{\IP^2}=\left(\re^{\hat x} +\re^{\hat p}+\re^{-\hat x-\hat p}\right)^{-1}. \ee
An important achievement of \cite{ghm, cgm2} is that one can compute \eqref{gsd} explicitly in term of (refined) Gopakumar-Vafa (GV) invariants of X. These topological invariants determine  \eqref{gsd} through the so--called  topological string grand potential $J_X({\boldsymbol \mu},\hbar)$  
which was  first introduced in \cite{mp} and then further studied in \cite{hmo, hmo2,hmo3,cm,hmmo,hoo,ghm,gkmr,hatsuda}.
More precisely we define  
\be\label{Jtot} \ba \mathsf{J}_X( {\boldsymbol \mu},\hbar)=&{1\over 12 \pi \hbar} a_{IJK} t_I(\hbar) t_J(\hbar) t_K(\hbar)  + {2\pi \over \hbar}b_I t_I(\hbar)+ {\hbar \over 2\pi  }b_I^{\rm NS} t_I(\hbar)+F^{\rm GV}({4 \pi^2 \over \ri \hbar},{2\pi \over \hbar}{\bf t}(\hbar)+\ri \pi B) \\
&+ \mathsf{J}_{b}( {\boldsymbol \mu},\hbar)+\mathsf{J}_{c}( {\boldsymbol \mu},\hbar) + A(\hbar).\\
\ea\ee
 We use
\be \boldsymbol{\kappa}=\re^{\boldsymbol{\mu}} \ee
and  
\be F^{\rm GV}(g_s,{\bf t}(\hbar))=\sum_{g\ge 0} \sum_{ \bf d} \sum_{w=1}^\infty {1\over w} n_g^{ {\bf d}} \left(2 \sin { w \ri  g_s \over 2} \right)^{2g-2} \re^{-w {\bf d} \cdot {\bf t}(\hbar)},
\ee
where $n_g^{ {\bf d}} $ are the standard GV invariants of $X$.  This quantity is sometimes called the instanton part of the  topological string free energy whose full expression reads 
 \be \label{Fst} F(g_s,{\bf t}(\hbar))=\sum_{g\geq 0}g_s^{2g-2}F_g({\bf t}(\hbar))={1\over 6 (\ri g_s)^2} a_{IJK} t_I(\hbar) t_J(\hbar) t_K(\hbar)  +b_I t_I(\hbar)  +F^{\rm GV}(g_s,{\bf t}(\hbar)).\ee
 We denote by  $F_g$ the genus g free energies of standard topological strings \footnote{Notice that in our notation $F_{2g}$  have a - sign w.r.t.~the notation in \cite{ghm, cgm2}.}.
 
The refined invariants $N^{{\bf d}}_{j_L, j_R} $ appear through  the functions $\mathsf{J}_{c}$ and $\mathsf{J}_{b}$. These are defined as
\be \mathsf{J}_{b}( {\boldsymbol \mu},\hbar)=\sum_{{\boldsymbol \ell} \geq {\bf 1}} ( \boldsymbol \ell \cdot {\bf t} ) b_{\boldsymbol \ell}\re^{- \boldsymbol \ell \cdot {\bf t}(\hbar) },\quad \mathsf{J}_{c}( {\boldsymbol \mu},\hbar)=\sum_{{\boldsymbol \ell} \geq {\bf 1}} c_{\boldsymbol \ell}\re^{- \boldsymbol \ell \cdot {\bf t}(\hbar) }, \ee
  with
  \be \label{bccoef}\ba   b_{\boldsymbol \ell}=&-\frac{1}{4 \pi}\sum_{j_L,j_R}\sum_{\boldsymbol \ell=\boldsymbol d  w}\sum_{\boldsymbol d}N^{\boldsymbol d}_{j_L,j_R}
\frac{\sin\frac{\hbar w}{2}(2j_L+1)\sin\frac{\hbar w}{2}(2j_R+1)}{w^2\sin^3\frac{\hbar w}{2}}.\\
  c_{\boldsymbol \ell}=&-\hbar^2{\partial \over \partial \hbar}\left({ b_{\boldsymbol \ell}\over \hbar} \right).\\
  \ea \ee%
 Generically, the geometry of the mirror curve to toric CYs is parametrized by two set of variables: the "true" moduli \eqref{kappa} and the mass parameters $m_i$. In this paper we set these parameters to  the particular value which guarantee the vanishing of the corresponding algebraic mirror map. As a consequence one has a particularly simple relation between the $  b_{\boldsymbol \ell}$'s and $  c_{\boldsymbol \ell}$'s coefficients given in \eqref{bccoef} (see \cite{ghm, gkmr} for more details).  
 
  Notice that $\mathsf{J}_{b}$ is closely related to the NS free energy 
  \be  \label{Fns}  \ba \hbar^{-1} F^{\rm NS}(\hbar, {\bf t}(\hbar))=\sum_{g\geq 0}\hbar^{2g-2}F_g^{\rm NS}({\bf t}(\hbar))&=- {1\over 6 \hbar^2} a_{IJK} t_I(\hbar) t_J (\hbar)t_K(\hbar)-b^{\rm NS}_I t_I 
  \\ &-\sum_{j_L, j_R} \sum_{w, {\bf d} } 
N^{{\bf d}}_{j_L, j_R}  \frac{\sin\frac{  \ri \hbar w}{2}(2j_L+1)\sin\frac{   \hbar w}{2}(2j_R+1)}{2  \hbar  w^2 \sin^3\frac{  \hbar w}{2 }} \re^{-w {\bf d}\cdot{\bf  t}( \hbar)}.  \ea\ee
The parameter  $B$  appearing in \eqref{Jtot} is a geometrical parameter which is related to the canonical class of the geometry X, we refer to it as B field \cite{hmmo}. The function $A(\hbar)$ in  \eqref{Jtot}  is the so-called constant map contribution \cite{bcov} and it is necessary  to guarantee the  correct normalization of the spectral determinant, namely
\be \Xi_X( {\boldsymbol 0},\hbar)=1.\ee
The conjecture of \cite{ghm, cgm2} states that
\be \label{sdtot}\Xi_X( {\boldsymbol \kappa}, \hbar)= \sum_{ {\bf n} \in \IZ^{g_\Sigma}} \exp \left( \mathsf{J}_{X}(\boldsymbol{\mu}+2 \pi \ri  {\bf n},  \hbar) \right). \ee
Moreover 
\be \label{xitoN}\Xi_X( {\boldsymbol \kappa}, \hbar)=\sum_{{ N_1}\geq { 0}} \cdots \sum_{{ N_{g_\Sigma}}\geq { 0}} Z({\bf{N}}, \hbar)\kappa_1^{N_1} \cdots \kappa_{g_\Sigma}^{N_{g_\Sigma}}\ee
where $Z({\bf N}, \hbar)$ defines the non--perturbative partition function of topological strings on $X$. As explained in \cite{ghm} it corresponds to the partition function of an ideal Fermi gas.
This conjecture and its consequences have been further tested in \cite{kama,mz,kmz,ho2,gkmr,oz,hw,wzh}. In \cite{hm,fhma} this conjecture has been used to obtain exact quantization conditions for the integrable systems of Goncharov and Kenyon and for the relativistic Toda lattice. In \cite{bgt}, by using recent developments in the context of Painlev\'e equations \cite{gil,bes,ilt}, a proof of the conjecture in a limiting case was provided.

The sum in \eqref{sdtot} can be implemented  formally and we  write
\be \label{sdtot22}\Xi_X( {\boldsymbol \kappa}, \hbar)= \exp \left( \mathsf{J}_{X}(\boldsymbol{\mu}+2 \pi \ri  {\bf n},  \hbar) \right) \Theta_X({\boldsymbol \kappa}, \hbar), \ee
where \be \Theta_X({\boldsymbol \kappa}, \hbar)=\sum_{ {\bf n} \in \IZ^{g_\Sigma}} \exp \left[ \mathsf{J}_{X}(\boldsymbol{\mu}+2 \pi \ri  {\bf n},  \hbar) -\mathsf{J}_{X}(\boldsymbol{\mu},  \hbar)  \right] \ee
defines the  {\it{quantum theta function}} whose zeros determine the spectrum of the operators \eqref{rhoop}. Moreover, according to the conjecture, its inclusion cures the non--analyticity of  the grand potential in such a way that the resulting spectral determinant is an analytic function in $ \boldsymbol \kappa$.
As explained in \cite{ghm, cgm2}, the grand potential \eqref{Jtot}  leads to an explicit expression  for the quantum theta function in terms of (refined) GV invariants valid near the large radius region of the moduli space  and, for real values of $\hbar$, it has good convergence properties. However away from this region very little is known. There are few exceptions to this,  among them we have  the self-dual point with  $\hbar=2\pi$. For this value the quantum theta function becomes a conventional theta function an 
one can write a closed form expression for \eqref{sdtot22} at any point of the moduli space in terms of hypergeometric and theta functions\footnote
{There are also few other values of $\hbar$ for which it has been possible  to resum the  Gopakumar--Vafa resummation and write down a closed form expression for \eqref{sdtot22}. 
This however requires some guess work since we have contributions from all genera and so far this resummation has been performed only in few cases \cite{cgm,ghmabjm, oz}.}.

In this paper we  explore the other regions of the moduli space away from the large radius point by performing an expansion around this special value of $\hbar=2\pi$.
As first pointed out in a related context \cite{cgm},  it is straightforward to see  that for this value of $\hbar$ only genus zero and genus one free energies
\be F_0({\bf t}(\hbar)), F_1({\bf t}(\hbar)), F_1^{\rm NS}({\bf t}(\hbar))\ee
contribute to \eqref{Fst} and \eqref{Fns}.
For  toric Calabi-Yaus these are typically known in closed form in term of Meijer and hypergeometric functions. Hence we can write down a closed form expression for $\Xi_X( {\boldsymbol \kappa}, 2\pi)$ at any point of the moduli space in term of these special functions. Let us define
 \be  \ba
 {F}_g=F_g({\bf t}(0)), &\quad  {F}_g^{\rm NS}=F_g^{\rm NS}({\bf t}(0)),
 \\
  \widehat{F}_g=F_g({\bf t}(2\pi)), &\quad  \widehat{F}_g^{\rm NS}=F_g^{\rm NS}({\bf t}(2\pi)). \ea \ee
In \cite{ghm,cgm2} it was argued that $ {\bf t}(2\pi) $
 can be obtained from 
 $ {\bf t}(0)$
 by switching the sign\footnote{This depends on the B field as explained in \cite{ghm,cgm2}.} in some of the complex deformations parameter describing the mirror geometry to X.
 Therefore, under a change of symplectic frame 
  \be {\bf{t}}(2\pi), \quad  \widehat{F}_g, \quad  \widehat{F}_g^{\rm NS},\ee
have the same tranformation properties of  \be  {\bf{t}}(0), \quad F_g, \quad F_g^{\rm NS}.\ee 

It was found in \cite{ghm,cgm2} that, when $\hbar=2\pi$, the spectral determinant takes a particularly simple form, namely
 \be \label{sd2pi} \Xi_X ( {\boldsymbol \kappa}, 2 \pi)= \exp\left[-{1\over 4 \pi^2} \widehat{F}_0 + \widehat F_1-\widehat F_1^{\rm NS}\right] \Theta_{0,\beta}( \widehat {\boldsymbol{\xi}},\widehat \tau),\ee
where \be \label{xitau}\ba & \widehat \xi_M= -{\mathcal{C}_{JM} \over 4 \pi^2} \left\{  {\partial^2 \widehat F_0 \over \partial t_J  \partial t_K} t_K  -  
{\partial \widehat F_0 \over \partial t_J } \right\}={1\over 4 \pi^2}\left( {\partial^2 \widehat F_0 \over \partial T_M  \partial T_K} T_K  -  
{\partial \widehat F_0 \over \partial T_M }\right), \qquad M=1, \cdots, g_\Sigma\\
 & \widehat \tau_{LM}= {1\over 2\pi \ri}\mathcal{C}_{JL} \mathcal{C}_{KM} {\partial^2 \widehat F_0 \over \partial t_J\partial t_K}= {1\over 2\pi \ri}\partial_{T_L}\partial_{T_M}\widehat F_0, \qquad L,M=1, \cdots, g_\Sigma.\ea\ee
 Notice that $\widehat \tau$  is modular parameter of the spectral curve \eqref{sf} therefore ${\rm Im}(\widehat \tau) >0$.
The constant $\beta$ in \eqref{sd2pi} can be computed explicitly in term of $b_I, b_J^{\rm NS}, a_{IJK}$ \cite{ghm,gkmr}.
The theta function $ \Theta_{\alpha,\beta}$ is defined as \cite{em} 
\be \label{qt} \Theta_{0,\beta}(\widehat  {\boldsymbol \xi}, \widehat \tau)=\exp\left[{1\over 4 \pi^2}{\left({\boldsymbol T}\partial_{\boldsymbol T} {\widehat F_0}-{1\over 2}{\boldsymbol T}^2\partial_{\boldsymbol T}^2\widehat F_0\right)}\right]\vartheta \left[\begin{array}{cc} \boldsymbol{0}\\ \boldsymbol{\beta}
 \end{array}\right](\widehat  {\boldsymbol \xi}, \widehat \tau ) ,\ee
where 
\be \label{thetadef}  \vartheta \left[\begin{array}{cc} \boldsymbol{\alpha}\\ \boldsymbol{\beta}
 \end{array}\right] \left({\boldsymbol{v}}, \tau \right)= \sum_{ {\boldsymbol{n}} \in \IZ^2 }\exp \left[  ~^t ({\boldsymbol{n}}  + {\boldsymbol{\alpha}}) {\tau  \ri \pi}  
({\boldsymbol{n}}  + {\boldsymbol{\alpha}}) +2 \pi \ri ({\boldsymbol{v}}  + {\boldsymbol{\beta}})  \cdot\left( {\boldsymbol{n}} + {\boldsymbol{\alpha}} \right) \right].
\ee
Moreover the K\"ahler parameters in \eqref{xitau} are evaluated at $\hbar=2\pi$.

As first noticed in \cite{cgm}, the spectral determinant evaluated at $\hbar=2\pi$  is similar to the leading order of the background independent partition function proposed in \cite{bde, eynard, em}. However, there are two main differences. The first one consists in the fact that the theta function has now an oscillatory behavior with no need to impose additional constraints on the moduli as was required in \cite{bde, eynard, em}.  This  improvement  of the convergence properties is related to the fact that to compute the spectral determinant \eqref{sdtot} one has to sum over imaginary shifts of the moduli, whereas  in \cite{bde,em,eynard} the sum runs over real shifts.
Moreover the proposal of \cite{bde,em,eynard} is
defined by a formal $1/N$ 't Hooft like expansion \eqref{thl} which misses important non-perturbative effects  in $\re^{-N}$ that in our formalism are determined by the NS free energies.
 These differences become even more important at higher orders in the expansion \eqref{exp}. In particular the NS quantities do not simply factorize as in \eqref{thetadef} but they mix with the  unrefined  free energies and their derivatives. We work out some explicit example in section \ref{qtf} and in appendix \ref{apo3}.

Notice that  $F^{\rm GV}$, $\mathsf{J}_b$ and $\mathsf{J}_c$ in \eqref{Jtot} are ill defined for $\hbar=2\pi$ due to the presence of some poles. However when we sum them up these poles disappear and we are left with a well defined quantity: the grand potential $J_X$. This is the so-called HMO cancellation mechanism \cite{hmo2} and to guarantee  such a mechanism the NS free energies play a crucial role.

As we will see in the next section, the genus one free energy $F^{\rm NS}_1$ is modular invariant. Therefore it follows from  the computations of \cite{em},  that $\Xi( {\boldsymbol \kappa},2\pi)$ is  invariant under \eqref{cytra} up to a phase and change of characteristic in the theta function.
Let us review how this goes. By using\footnote{As in \eqref{xitau}  the K\"ahler parameters ${\bf T}$  are evaluated at $\hbar=2\pi$. }
\be \label{pt2b}\left(\begin{matrix} \hbar \partial_{{\bf T} }\widehat  F_0 \\ 2\pi \ri {\bf T} \end{matrix}\right) \to  \left(\begin{matrix} A & B \\ C & D \end{matrix}\right)  \left(\begin{matrix} \hbar \partial_{{\bf T }} \widehat F_0 \\ 2\pi \ri {\bf T}  \end{matrix}\right),\ee
it is straightforward to see that the combination
\be \label{m1}\widehat{F}_0 -{1\over 2}{\boldsymbol T}\partial_{\boldsymbol T} {\widehat F_0},\ee
is symplectic invariant. Similarly we have
\be \label{m2}\ba  - {N^2 \over 2}\left( {\boldsymbol T}\partial_{\boldsymbol T} {\widehat F_0}-{\boldsymbol T}^2\partial_{\boldsymbol T}^2\widehat F_0\right) \to& 
 - {N^2\over 2}\left( {\boldsymbol T}\partial_{\boldsymbol T} {\widehat F_0}-{\boldsymbol T}^2\partial_{\boldsymbol T}^2\widehat F_0\right) +\ri \pi \widehat {\boldsymbol{\xi} }\widehat{\mathcal K} \widehat {\boldsymbol{\xi}},\\
 \widehat {\boldsymbol{\xi} } \to & \widehat {\boldsymbol{\xi} } \left( C \widehat \tau +D\right)^{-1},\\
 \widehat \tau \to & \left( A \widehat \tau +B\right)\left( C \widehat \tau +D\right)^{-1},
\ea\ee
where
\be \label{kappahat} \widehat{\mathcal K}= -C \left( C \widehat \tau +D\right)^{-1}.\ee
In particular one can show that, up to an overall phase factor, we have \cite{em}
\be   \label{m3}\re^{ \widehat F_1}\vartheta \left[\begin{array}{cc} \boldsymbol{0}\\ \boldsymbol{\beta}
 \end{array}\right](\widehat  {\boldsymbol \xi}, \widehat \tau ) \to    \re^{ \widehat F_1}\re^{-\pi \ri  \widehat {\boldsymbol{\xi} }\widehat {\mathcal K} \widehat {\boldsymbol{\xi}} }\vartheta \left[\begin{array}{cc} \boldsymbol{\gamma}\\ \boldsymbol{\delta}
 \end{array}\right](\widehat  {\boldsymbol \xi}, \widehat \tau ),\ee
 with
\be \ba 
&C \boldsymbol \delta = D \boldsymbol \gamma  +{1\over 2}(CD^T)_d, \\
&\boldsymbol \beta=A\boldsymbol \delta -B \boldsymbol \gamma +{1\over 2}(AB^T)_d,
\ea\ee
where $(M)_d$ is a  vector whose components are the diagonal element of the matrix $M$.
Hence from \eqref{m1},\eqref{m2} and  \eqref{m3} it follows that  $\Xi( {\boldsymbol \kappa},2\pi)$ is  invariant under a change of symplectic frame \eqref{cytra}, up to a phase factor and a change of characteristic in the theta function.

Notice that the $\widehat \tau$ matrix in \eqref{xitau} reads
\be \widehat \tau_{IJ}=\partial_{\epsilon_I}\partial_{\epsilon_J}\widehat F_0.\ee
Similarly we define
\be  \tau_{IJ}=\partial_{\epsilon_I}\partial_{\epsilon_J} F_0,\ee
where 
\be F_0=F_0({\bf t}(0)).\ee
According to
\eqref{cltra} we have
\be \tau \to\left( {A \tau+B}\right)\left(C\tau +D\right)^{-1}.\ee

\section{ Change of symplectic frame }\label{mns}
We are interested in studying the spectral determinant at a generic point of the moduli space. As explained in \cite{we,abk}   different points in the moduli space correspond to different  choices of  symplectic frames \eqref{cytra}.
Therefore we have to study the symplectic properties of the quantities which determine the spectral determinant $\Xi( {\boldsymbol \kappa},\hbar)$, in particular the standard and NS free energies \eqref{Fst}, \eqref{Fns}.  For the standard free energy \eqref{Fst} these properties have been worked out in \cite{abk,emo} where it was shown than these can be formulated in a compact way thanks to the wave function behavior of the unrefined partition function.  A discussion of the modular properties of the NS free energies has been done in \cite{hk,huangNS}. In the following we organize them by using diagrammatic rules in a wave function like behavior which is controlled by the same action of the unrefined theory.

\subsection{The Nekrasov--Shatashvili free energies } \label{sdr}
In this section we study how the NS  free energies transform under \eqref{cytra}. For that we need  to integrate $\Pi_{B_I}(\hbar,  {\boldsymbol \kappa})$ by carefully  taking into account the $\hbar$ dependence in the quantum A period ${\boldsymbol \epsilon} (\hbar, {\boldsymbol \kappa})$. 
We will use
\be \label{hexp}\ba F^{\rm NS}_g=& F^{\rm NS}_g({\bf t}(0)),\\
\epsilon_I(\hbar)=&\sum_{n \geq 0} \hbar^{2 n}\epsilon^{(n)}_I.
\ea\ee
Then we have ($\partial_I= \partial_{ \epsilon_I} $)
\be \label{exo}\ba \hbar \partial_{ I} F^{\rm NS}(\hbar,{\bf t}(\hbar))= &({  \partial_I} F^{\rm NS}_0+ \hbar^2\left( { \partial_I} F^{\rm NS}_1+{ \epsilon_J}^{(1)}\tau_{JI}\right)\\ 
& +\hbar^4 \left( { \epsilon_J}^{(2)}\tau_{IJ} +{ \epsilon_J}^{(1)}{ \epsilon_K}^{(1)}{1\over 2} \partial_K \partial_J \partial_IF_0 +{\epsilon_K}^{(1)}\partial_K\partial_IF_1^{\rm NS}+\partial_I  F_2^{\rm NS}\right)+\mathcal{O}\left(\hbar^6\right)
\ea \ee
where we denote by $ \partial_IF^{\rm NS}_g$ the derivative w.r.t.~the classical period ${ \epsilon_I}^{(0)}$ evaluated at $\hbar=0$. Since the index structure of \eqref{exo} is clear we will simply denote it by
\be \ba \hbar \partial F^{\rm NS}= &\partial F^{\rm NS}_0+ \hbar^2\left( \partial F^{\rm NS}_1+\epsilon^{(1)}\tau\right)\\ 
& +\hbar^4 \left( \epsilon^{(2)}\tau +\left(\epsilon^{(1)}\right)^2{1\over 2} \partial^3F_0 +\epsilon^{(1)}\partial^2F_1^{\rm NS}+\partial F_2^{\rm NS}\right)+\mathcal{O}\left(\hbar^6\right).
\ea \ee
By carefully computing the $\hbar$ expansion on the l.h.s~and on the r.h.s.~of \eqref{pt}  and by performing the integration over $\epsilon$ we observe that the modular properties of $F_g^{\rm NS}$ can be encoded by  using diagrammatic rules similar to the ones used for standard free energy
\cite{eo, emo}.

Let us represent  $\partial^s F_g^{\rm NS}$ by a Riemann surface of genus $g$ with $s$ punctures. We observe that, under the modular transformation \eqref{cytra}, the free energy $F_g^{\rm NS}$ transforms
into itself plus the sum of all possible connected (via propagators) Riemann surfaces of total genus $g$  fulfilling
\be \label{rule} m-n=-1, \qquad \sum_i g_i=g, \quad g_i<g , \quad \ee
where $m$ is the number of propagators and $n$ the number of surfaces connected by  propagators. As in \cite{eo,emo} we require at least 3 punctures on a surface of genus zero and  the propagators are given by
\be\label{taukappa}
\ba
 & {\mathcal K}=- C \left( C \tau +D\right)^{-1}. \ea \ee
\begin{center}
 \begin{figure} \begin{center}
 {\includegraphics[scale=0.45]{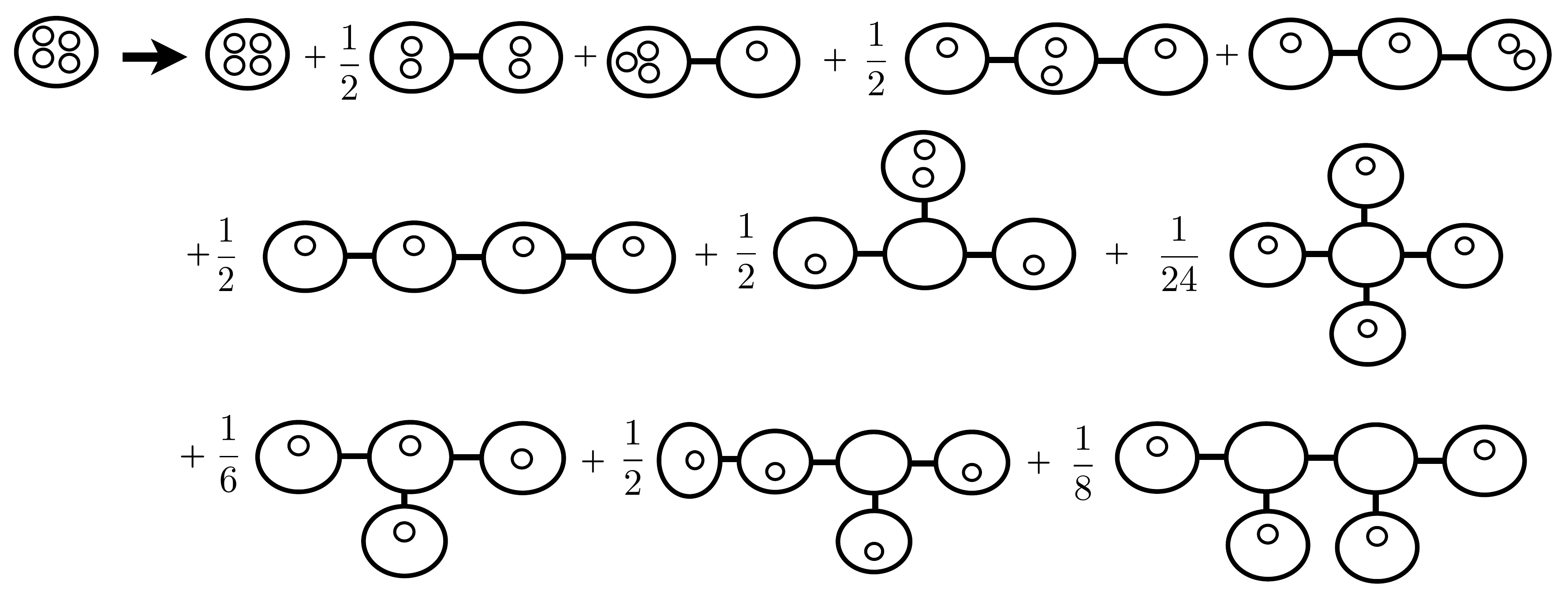}}
\caption{  The modular transformation of $F_4^{\rm NS}$. }
 \label{picturemt}
  \end{center}
\end{figure}  \end{center}
Moreover  the weight $1/s$ of each graph is obtained simply by counting the multiplicity of each surface inside the graph  as in the unrefined case.
This is shown for instance on  \figref{picturemt} for $F_4^{\rm NS}$. 
\raggedbottom

Let us look at some concrete examples.   Starting from \eqref{pt}, one can show that
  \be\label{qpt}\ba
 &\partial F_{0} \to  A \partial F_{0} +B\epsilon^{(0)},\\
 &\partial  F_{1} ^{\rm NS}\to  \left( C \tau +D\right)^{-1}\partial F_{1}^{\rm NS},  \\
 &\partial  F_{2} ^{\rm NS}\to  \left( C \tau +D\right)^{-1}  \left(\partial F_{2} ^{\rm NS}+  {\mathcal K} \partial F_{1}^{\rm NS} \partial ^2 F_{1}^{\rm NS} \right. \\
 & \qquad \qquad \qquad \qquad  \quad\qquad \left.+ {1\over 2} \partial ^3 F_{0} \left(\partial F_{1}^{\rm NS}\right)^2  {\mathcal K}^2   \right),\\
   \ea\ee
 where we used that $\partial ^3 F_{0} $ behaves as a weak Jacobi form of weight -3 under \eqref{cytra} as shown for instance in \cite{abk}. 
 By integrating \eqref{qpt} we obtain that $F_1^{\rm NS}$ is modular invariant while
 \be \label{F2mt}
 F_2^{\rm NS} \to F_2^{\rm NS} + {1\over 2} {\mathcal K}\left(\partial  F_{1}^{\rm NS}\right)^2.
\ee
Similarly one can show that
\be \label{MT34} \ba  F_3^{\rm NS}&\to F_3^{\rm NS}+\frac{1}{6}  {\mathcal K} ^3 \partial^3 F_0( \partial F_1^{\rm NS})^3+ {\mathcal K}  \partial F_1^{\rm NS} \partial F_2^{\rm NS} +\frac{1}{2}  {\mathcal K} ^2( \partial F_1^{\rm NS} )^2 \partial^2 F_1^{\rm NS},\\
F_4^{\rm NS} &\to F_4^{\rm NS}+\frac{1}{2}  {\mathcal K} ( \partial F_2^{\rm NS})^2 + {\mathcal K}  \partial F_1^{\rm NS} \partial F_3^{\rm NS}+\frac{1}{2}  {\mathcal K} ^2 (\partial F_1^{\rm NS})^2 \partial^2 F_2^{\rm NS}+ {\mathcal K} ^2 \partial F_1^{\rm NS} \partial^2 F_1^{\rm NS} \partial F_2^{\rm NS}\\
&+\frac{1}{2}  {\mathcal K} ^3 (\partial F_1^{\rm NS})^2 (\partial^2 F_1^{\rm NS})^2
+\frac{1}{6}  {\mathcal K} ^3 \partial^3 F_1^{\rm NS} (\partial F_1^{\rm NS})^3+\frac{1}{24}  {\mathcal K} ^4 \partial^4 F_0^{\rm NS} (\partial F_1^{\rm NS})^4
\\&+\frac{1}{2}  {\mathcal K} ^3 \partial^3 F_0^{\rm NS} (\partial F_1^{\rm NS})^2 \partial F_2^{\rm NS}+\frac{1}{2}  {\mathcal K} ^4 \partial^3 F_0( \partial F_1^{\rm NS})^3 \partial^2 F_1^{\rm NS}+\frac{1}{8}  {\mathcal K}^5 \partial^3 F_0^2 (\partial F_1^{\rm NS})^4,\\
 \ea\ee
 \be \ba \label{MT5}
 F_5^{\rm NS}  
 \to  &F_5^{\rm NS}+ \frac{1}{2} \mathcal{K} ^2\partial^2 F_1^{\rm NS}\lt \partial F_2^{\rm NS}\rt^2+\mathcal{K} \partial F_1^{\rm NS}  \partial F_4^{\rm NS}+\mathcal{K} ^3\partial F_1^{\rm NS}\left(\partial^2 F_1^{\rm NS}\right)^2 \partial F_2^{\rm NS}\\
&+\mathcal{K} ^2\partial F_1^{\rm NS}\partial^2 F_1^{\rm NS} \ \partial F_3^{\rm NS}+\mathcal{K}  \partial F_2^{\rm NS}  \partial F_3^{\rm NS} \\
&+\frac{1}{2} \mathcal{K} ^3 \partial^3 F_0\partial F_1^{\rm NS}\lt \partial F_2^{\rm NS}\rt^2+\mathcal{K} ^2\partial F_1^{\rm NS} \partial F_2^{\rm NS} \partial^2 F_2^{\rm NS}+\frac{1}{2} \mathcal{K} ^2\lt\partial F_1^{\rm NS}\rt^2 \partial^2 F_3^{\rm NS}\\
& +\mathcal{K} ^3\lt\partial F_1^{\rm NS}\rt^2\partial^2 F_1^{\rm NS} \partial^2 F_2^{\rm NS}+\frac{1}{2} \mathcal{K} ^4\lt \partial F_1^{\rm NS}\rt^2\left(\partial^2 F_1^{\rm NS}\right)^3 +\frac{1}{2} \mathcal{K} ^4 \partial^3 F_0\lt\partial F_1^{\rm NS}\rt^3 \partial^2 F_2^{\rm NS}
\\&+\frac{1}{2} \mathcal{K} ^3 \partial^3 F_0\left(\partial F_1^{\rm NS}\right)^2 \ \partial F_3^{\rm NS}+\frac{3}{2} \mathcal{K} ^4 \partial^3 F_0\left(\partial F_1^{\rm NS}\right)^2\partial^2 F_1^{\rm NS} \partial F_2^{\rm NS} \\
&+\mathcal{K} ^5 \partial^3 F_0\lt\partial F_1^{\rm NS}\rt^3\lt\partial^2 F_1^{\rm NS}\rt^2+\frac{1}{2} \mathcal{K} ^5 \partial^3 F_0^2\left(\partial F_1^{\rm NS}\right)^3 \partial F_2^{\rm NS}\\
&+\frac{1}{8} \mathcal{K} ^7 \partial^3 F_0^3\lt\partial F_1^{\rm NS}\rt^5+\frac{5}{8} \mathcal{K} ^6 \partial^3 F_0^2\lt\partial F_1^{\rm NS}\rt^4\partial^2 F_1^{\rm NS} +\frac{1}{2} \mathcal{K} ^3 \partial^3 F_1^{\rm NS}\lt\partial F_1^{\rm NS}\rt^2 \partial F_2^{\rm NS}\\
&+\frac{1}{6} \mathcal{K} ^4 \partial^4 F_0^{\rm NS}\lt\partial F_1^{\rm NS}\rt^3 \partial F_2^{\rm NS}+\frac{1}{4} \mathcal{K} ^5 \partial^3 F_0 \partial^3 F_1^{\rm NS}\lt\partial F_1^{\rm NS}\rt^4+\frac{1}{6} \mathcal{K} ^3 \partial^3 F_2^{\rm NS}\lt\partial F_1^{\rm NS}\rt^3\\
&+\frac{1}{2} \mathcal{K} ^4 \partial^3 F_1^{\rm NS}\lt\partial F_1^{\rm NS}\rt^3\partial^2 F_1^{\rm NS}+\frac{1}{120} \mathcal{K} ^5 \partial^5 F_0^{\rm NS}\lt\partial F_1^{\rm NS}\rt^5+\frac{1}{6} \mathcal{K} ^5 \partial^4 F_0^{\rm NS}\lt\partial F_1^{\rm NS}\rt^4\partial^2 F_1^{\rm NS}\\
&+\frac{1}{12} \mathcal{K} ^6 \partial^3 F_0 \partial^4 F_0\lt\partial F_1^{\rm NS}\rt^5+\frac{1}{24} \mathcal{K} ^4\partial^4 F_1^{\rm NS}\lt\partial F_1^{\rm NS}\rt^4 .\\
 \ea\ee
 \begin{center}
 \begin{figure} \begin{center}
 {\includegraphics[scale=0.3]{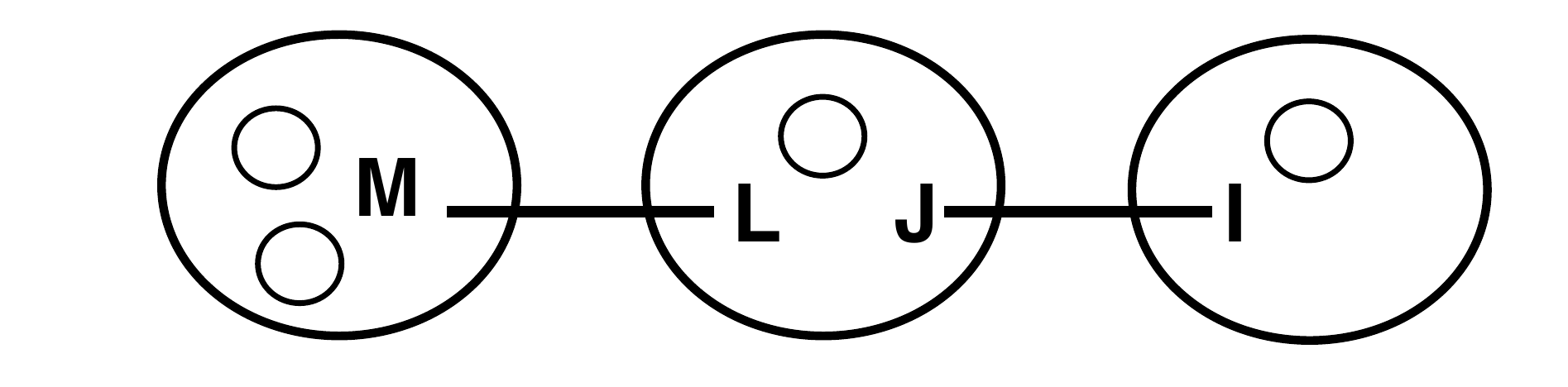}\\ \includegraphics[scale=0.3]{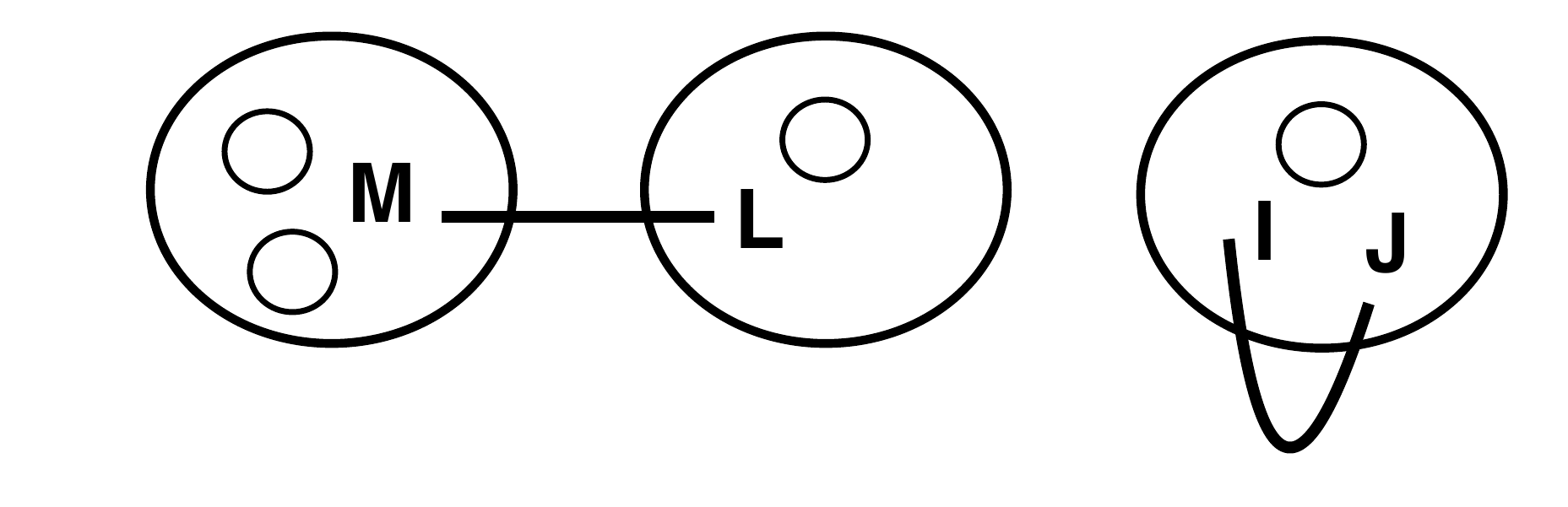}}
\caption{A graphical representation of \eqref{indexok} (top), and of \eqref{indexno} (bottom). }
 \label{indexs}
  \end{center}
\end{figure}  \end{center}
 The above transformations have been worked out by an explicit computation from \eqref{pt}, however it is easy to check that they can be derived by using the diagrammatic rules explained around equation \eqref{rule}. Similarly we have explicitly check  these diagrammatic rules  also for higher $F_g$. 
 \raggedbottom

The  structure of the indices in \eqref{MT34} and \eqref{MT5} is always such that the corresponding diagram is connected. For instance let us consider the term
\be  {\mathcal K} ^2 \partial F_1^{\rm NS} \partial^2 F_1^{\rm NS} \partial F_2^{\rm NS}\ee
appearing in the modular transformation of $F_4$ in \eqref{MT34}. The correct  structure of the indices is
\be \label{indexok} {\mathcal K}_{IJ} {\mathcal K}_{LM} \partial_I F_1^{\rm NS} \partial_J\partial_L F_1^{\rm NS} \partial_M F_2^{\rm NS}.\ee
This is represented by the diagram in the top of  \figref{indexs}. As an example let us consider the  following index structure
\be \label{indexno} {\mathcal K}_{IJ} {\mathcal K}_{LM} \partial_L F_1^{\rm NS} \partial_J\partial_IF_1^{\rm NS} \partial_M F_2^{\rm NS}.\ee
This is represented by the diagram in the bottom of  \figref{indexs}. This diagram is disconnected, and \eqref{indexno} has indeed the wrong index structure.
It could be that there are two "connected" index structures for a single term. This is the case for instance for the term
\be\label{egF2} { 5\over 24}  {\mathcal K}^3 \partial^3 \widehat F_0^2 \ee
appearing in the modular transformation of $F_2$ \eqref{F2tras}. In this case one has to consider all the connected index structures with the corresponding symmetry factor. For instance
\eqref{egF2} reads
\be  \ba & {1\over 12}{\mathcal K}_{IJ}  {\mathcal K}_{LM}  {\mathcal K}_{PQ} \partial_{I} \partial_{L}\partial_{P}\widehat F_0 \partial_{J} \partial_{M}\partial_{Q}\widehat F_0 +\\
& {1\over 8}{\mathcal K}_{IJ}  {\mathcal K}_{LM}  {\mathcal K}_{PQ} \partial_{I} \partial_{J}\partial_{P}\widehat F_0 \partial_{L} \partial_{M}\partial_{Q}\widehat F_0.
\ea\ee
 
\subsection{The quantum K\"ahler parameters}\label{sepsi}
\begin{center}
 \begin{figure} \begin{center}
 {\includegraphics[scale=0.45]{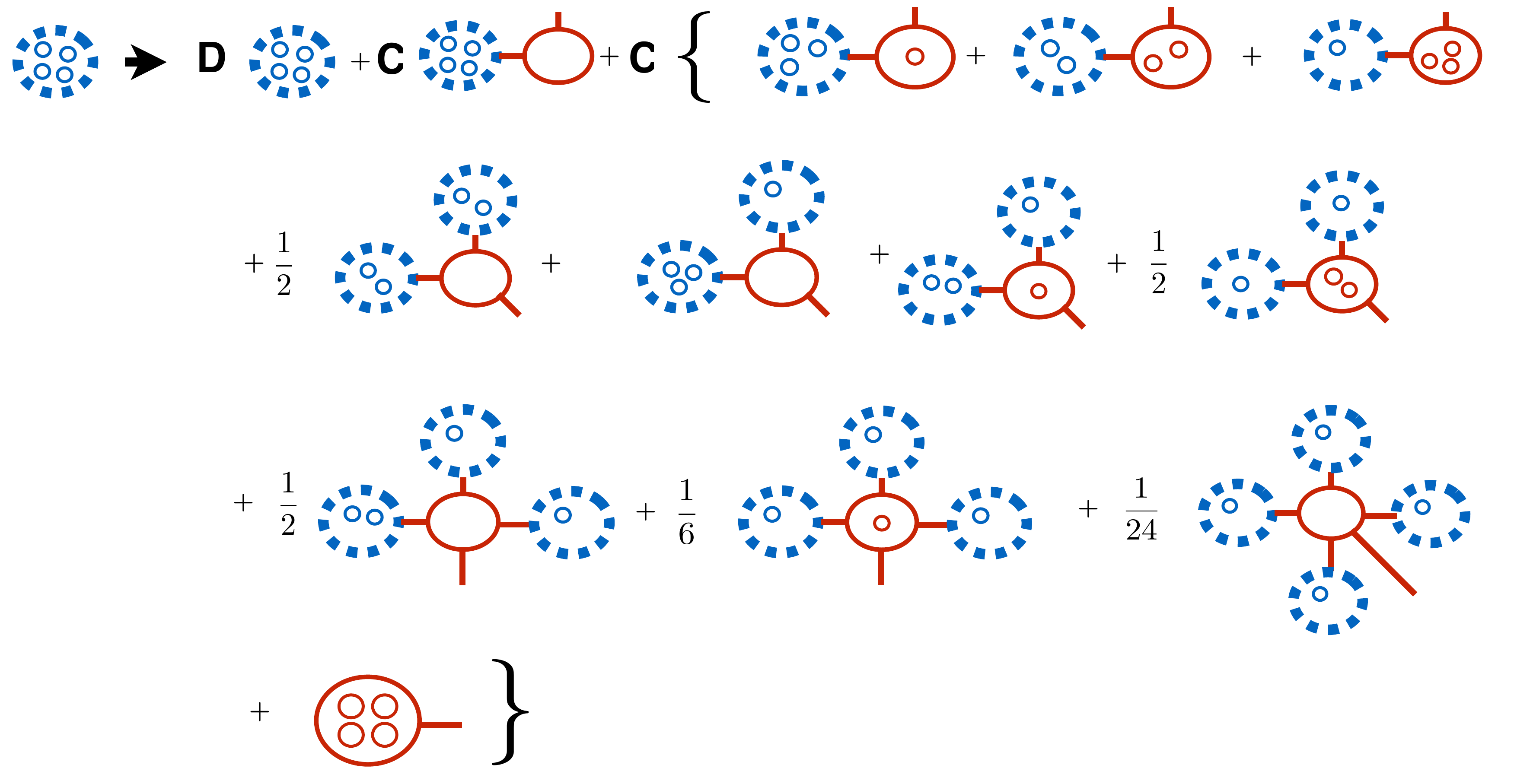}}
\caption{  The modular transformation of $\epsilon^{(4)}$. The blue dashed  surfaces represent the $n$ "satellites" and the red ones represent $\partial^{n+1}F_g^{\rm NS}$ }
 \label{epkmt}
  \end{center}
\end{figure}  \end{center}
\raggedbottom
Likewise, we notice that  the transformation properties of the ${\rm n }^{\rm th}$ component of the  A period, $\epsilon^{(n)}$, can be encoded by using
 diagrammatic rules. By doing an explicit computation one can show that \eqref{pt} leads to
\be\label{ept}\ba
  & \epsilon^{(0)} \to C\partial F_{0} +D\epsilon^{(0)} ,  \\
 & \epsilon^{(1)} \to C \left(\partial  F_{1}^{\rm NS}+ \epsilon^{(1)}\tau\right) +D\epsilon^{(1)}, \\
  & \epsilon^{(2)} \to C \left(\partial  F_{2} ^{\rm NS}+\epsilon^{(1)}\partial ^2 F_{1} ^{\rm NS}+{1\over 2} (\epsilon^{(1)} )^2\partial ^3 F_{0} + \epsilon^{(2)}\tau \right) +D\epsilon^{(2)},\\
& \epsilon^{(3)} \to  D \epsilon^{(3)}+C \left(\frac{1}{6}(\epsilon^{(1)})^3 \partial^4 F_0+\epsilon^{(1)}\epsilon^{(2)} \partial^3 F_0 \right. \\
 &\qquad \quad \left.+\epsilon^{(3)}\partial^2 F_0+\frac{1}{2}(\epsilon^{(1)})^2 \partial^3 F_1^{\rm NS}+\epsilon^{(2)}\ \partial^2 F_1^{\rm NS}+\epsilon^{(1)}\partial^2F_2^{\rm NS}+\partial F_3^{\rm NS}\right),\\
& \epsilon^{(4)} \to  D \epsilon^{(4)}+C \left( \partial F_4^{\rm NS}+ \epsilon^{(4)} \partial^2 F_0+\epsilon^{(3)}\partial^2 F_1^{\rm NS}+\epsilon^{(2)}\partial^2 F_2^{\rm NS}
 +\epsilon^{(1)}\partial^2 F_3^{\rm NS}+{1\over 2}(\epsilon^{(2)})^2 \partial^3 F_0\right. \\
 &\qquad \quad   \epsilon^{(1)}\epsilon^{(3)} \partial^3 F_0+\epsilon^{(1)}\epsilon^{(2)} \partial^3 F_1^{\rm NS} +{1\over 2}(\epsilon^{(1)})^2 \partial^3 F_2^{\rm NS}  
 +{1\over 2}(\epsilon^{(1)})^2 \epsilon^{(2)}\partial^4 F_0+{1\over 6}(\epsilon^{(1)})^3\partial^4 F_1^{\rm NS}
 \\ &\qquad \quad  \left.+{1\over 24}(\epsilon^{(1)})^4\partial^5 F_0
 \right).
 \ea \ee
We encode the above transformation by saying that
\be \epsilon^{(k)} \to  \left(D+C \tau\right)\epsilon^{(k)}+C \mathcal{G}(k), \ee
where $\mathcal{G}(k)$ can be represented as a sum of diagrams. Each of these diagrams is made of one Riemann surface of genus $g$ with one leg sticking out and $n\geq 0$ "satellites" of genus $k > g_i\geq 1$ ( representing $\epsilon^{(g_i)}$ ) such that
\be \sum_{i=1}^n g_i+g= k.\ee
This is shown on  \figref{epkmt}  for $k=4$. The multiplicity is computed in the standard way. 

In appendix \ref{app2} we check by an explicit computation the rules \eqref{ept} in the case of the local $\IP^2$ geometry.  The comparison between our results \eqref{ept} and the direct computation leads to interesting identities, as for instance \eqref{fid},\eqref{sid}, \eqref{tid}, which can be checked explicitly. \newline

\subsection{The wave function and the holomorphic anomaly} \label{haes}
 \raggedbottom
We observe that the  diagrammatic rules described in  section \eqref{sdr} can be encoded in some kind of wave function behavior similar to the one for the standard free energies \cite{emo, abk}.
 Let us define \cite{emo}
 \be \label{emoaction}
\hbar^2 S(\epsilon,\eta,\hbar,\mathcal{K})=-{1\over 2}(\eta-\epsilon) {\mathcal K}^{-1}(\eta-\epsilon)-(\eta-\epsilon)\partial_\epsilon F_0(\epsilon)-{1\over 2}(\eta-\epsilon) \partial_\epsilon^2F_0(\epsilon)(\eta-\epsilon),\ee
where $ {\mathcal K}$ is defined in \eqref{taukappa} and $\epsilon$ denotes here the classical period \eqref{cl}.
Then  the previous transformation rules can be written as
\be \label{wf}\re^{\hbar^{-1}  F^{\rm NS}(\hbar,\epsilon, {\mathcal K})}= \left({1\over \hbar \sqrt{2\pi  \det {\mathcal K}}}\int {\rm d} \eta \re^{ S(\epsilon,\eta,\hbar,\mathcal{K})+\hbar^{-1} {F}^{\rm NS}(\hbar,\eta)}\right)\Big|_{ g  ~ \rm{fixed}}, \ee
where  by $\mid_{ g  ~ \rm{fixed}}$ we mean that at each order in the $\hbar^{2g-2}$ expansion we keep only diagrams with total genus $g$.
We use $F^{\rm NS}(\hbar,\epsilon, {\mathcal K})$ to note the NS free energy obtained after performing the modular transformation \eqref{cytra}. When ${\mathcal K}=0$ we simply note   $F^{\rm NS}(\hbar,\epsilon)$.
To implement the constraint $\mid_{ g  ~ \rm{fixed}}$, it is useful to introduce an extra variable $y$ and consider
\be \label{wfy} Z_y(\mathcal{K})=\left({1\over \hbar \sqrt{2\pi  \det {\mathcal K}}}\int {\rm d} \eta \re^{ y^{-1}S(\epsilon,\eta,\hbar,\mathcal{K})+y^{-1}\hbar^{-1} {F}^{\rm NS}(\hbar,\eta)}\right).\ee
The transformations rules for the $F_g^{\rm NS}$'s can then be computed by doing a saddle point analysis of the integral appearing on the r.h.s.~of \eqref{wf} or \eqref{wfy} .
It is easy to see that the saddle point for small $\hbar$ is given by \be \eta=\epsilon.\ee
We explicitly checked that  the saddle point analysis of \eqref{wf} agrees with \eqref{F2mt}, \eqref{MT34}, \eqref{MT5} which have been at first derived from \eqref{pt}.
When we compute the saddle point of \eqref{wfy}, at each order  in the $\hbar^{2g-2}$ expansion, the diagrams with total genus $g$ are multiplied by a factor of $y^{-1}$. Similarly  the diagrams with total genus $g'<g$ are multiplied by a positive power of $y$. 
For instance at order $\hbar^2$
the saddle point of \eqref{wfy} gives
\be  S_2(\mathcal{K}, y)= y^{-1} F_2^{\rm NS}+y^{-1} \frac{1}{2} {\mathcal K} (\partial F_1^{\rm NS})^2 +\frac{1}{2} {\mathcal K} \partial^2 F_1^{\rm NS}+y \frac{1}{8} {\mathcal K} ^2 \partial^4 F_0+\frac{1}{2} {\mathcal K} ^2 \partial^3 F_0 \partial F_1^{\rm NS}+ y \frac{5}{24} {\mathcal K}^3 \partial^3 F_0^2.\ee
It follows that 
\be F_2^{\rm NS}(\epsilon,  \mathcal{K})= \lim_{y \to 0} { y}S_2( \mathcal{K}, y)=F_2^{\rm NS}+\frac{1}{2}  \mathcal{K} (\partial F_1^{\rm NS})^2.\ee
Therefore we have a systematic way to compute the l.h.s.~of \eqref{wf} as:
\be \label{limit}\sum_{g\geq0} \hbar^{2g-2} F^{\rm NS}_g(\epsilon, \mathcal{K}) =\lim_{y\to 0} y \log Z_y(\mathcal{K}). \ee
Notice that if we set $y=1$  we recover the wave function behavior of the standard topological string free energy \cite{abk,emo}.

By following \cite{emo}, one can derive from \eqref{wfy} the holomorphic anomaly equation for the NS free energies and check that it reproduce the well known result \cite{hk,hkpk,huangNS}.  
Let us introduce a non--holomorphic dependence   in $\overline \epsilon_I$ by setting \cite{emo}
\be \mathcal{K}_0=\left(\overline \tau- \tau\right)^{-1}\ee
and define
\be \label{zdef}\ba Z=& \left({1\over \hbar \sqrt{2\pi }}\int {\rm d} \eta \re^{ y^{-1} S(\epsilon,\eta,\hbar,\mathcal{K}_0)+y^{-1}\hbar^{-1} {F}^{\rm NS}(\hbar,\eta)}\right), \\
 \tilde Z=&  \left({1\over \hbar \sqrt{2\pi  \det {\mathcal K}_0}}\int {\rm d} \eta \re^{ y^{-1} S(\epsilon,\eta,\hbar,\mathcal{K}_0)+ y^{-1}\hbar^{-1} {F}^{\rm NS}(\hbar,\eta)}\right)\re^{-y^{-1} \hbar^{-2} F_0 (\epsilon)-y^{-1} F_1^{\rm NS}(\epsilon)}, \\
 \widehat Z=&\left({1\over \hbar \sqrt{2\pi }}\int {\rm d} \eta \re^{ y^{-1} S(\epsilon,\eta,\hbar,\mathcal{K}_0)+y^{-1}\hbar^{-1} {F}^{\rm NS}(\hbar,\eta)}\right)\re^{-y^{-1} \hbar^{-2} F_0(\epsilon)}.\ea \ee
 Then we have
 \be {1 \over \tilde Z}\partial_{\overline K} \tilde Z=
 {1\over 2 } \left( \mathcal{K}_0^{IJ}\partial_{\overline K} \overline \tau_{IJ}-y^{-1}\hbar^{-2}\langle(\eta-\epsilon)^I (\eta-\epsilon)^J  \rangle \partial_{\overline K} \overline \tau_{IJ} \right)\ee
 where $\langle \cdot \rangle$ denotes the expectation value w.r.t.~the integral in the first line of \eqref{zdef}. Similarly one has
 \be \ba {1 \over \widehat Z}\partial_{ I} \widehat Z =& y^{-1}\hbar^{-2}  \langle \left( \overline \tau-\tau\right)_{IJ}(\eta-\epsilon)^J \rangle, \\
  {1 \over \widehat Z}\partial_{ I}\partial_{J} \widehat Z =&- \hbar^{-2} y^{-1}\left( \overline \tau-\tau\right)_{IJ}+\hbar^{-4}y^{-2}\left( \overline \tau-\tau\right)_{IM}\left( \overline \tau-\tau\right)_{JL} \langle (\eta-\epsilon)^M (\eta-\epsilon)^L \rangle \\
  & -\hbar^{-2} y^{-1}\partial_I \tau_{JM}\langle (\eta-\epsilon)^M \rangle.
 \ea \ee
 In particular
 \be \label{haeo} y \partial_{\overline K} \log \tilde Z={y \over 2 } \hbar^2 \overline C^{IJ}_{\overline K }\left( y \partial_{ I}\partial_{J} \log \widehat Z+y \partial_{ I}\log \widehat Z \partial_{J} \log \widehat Z + y\mathcal{K}_0^{LM}\partial_I\tau_{JM} \partial_{L} \log \widehat Z\right), \ee
 where
 \be  \overline C^{IJ}_{\overline K }= -  {\mathcal K}_0^{IM} {\partial^3 \overline F_0 \over \partial \overline \epsilon_N \partial \overline \epsilon_M \partial \overline \epsilon_K} {\mathcal K}_0^{NJ}.\ee
When $y=1$ \eqref{haeo} takes  the form of the  holomorphic anomaly equation of unrefined topological strings \cite{bcov, emo}. 
If instead  we take $y \to 0$ in \eqref{haeo}, by using \eqref{limit}, we recover the refined holomorphic anomaly  equation for the  NS free energies \cite{hk,hkpk,huangNS}
\be  \label{hans}   \partial_{ \overline I} F^{\rm NS}_g= {1\over 2} \overline C^{JK}_{\overline I } \sum_{h\geq 1}\partial_J F_n^{\rm NS}\partial_K F_{g-n}^{\rm NS}, \quad g >1 .\ee

\section{Symplectic invariance and the spectral determinant} \label{qtf}

We are now going to use the results of section \ref{mns} to investigate the modular properties of the spectral determinant \eqref{sdtot}.  As explained previously the strategy is to expand the spectral determinant around $\hbar=2\pi$:
\be  \label{series}\Xi ( {\boldsymbol \kappa},\hbar)=\sum_{n\geq 0}{1\over n!}{\rd^n \Xi ( {\boldsymbol \kappa}, 2\pi) \over \rd^n \hbar}(\hbar-2\pi)^n, \ee
where the leading order $n=0$ is given by \eqref{sd2pi}. We find that the $n^{\rm th}$ order in this expansion can be written in closed form in terms of $F_g$, $F_h^{\rm NS}$ with $g,h \leq n$ and derivatives of the  theta function \eqref{qt}. 
In the following we  work out the first few orders in the  expansion \eqref{series} and we  show modular invariance by and explicit computation. We will then argue that this is the case at each order in the $\hbar-2\pi$ expansion.  This is done by using the combinatorial  techniques developed in \cite{em} together with the rules worked out in section  \ref{mns}.

We  use the shortcut notation
\be  \Theta=  \Theta_{0,\beta}(\widehat {\boldsymbol{\xi}}, \widehat \tau),\ee
together with \be
\left( \partial_{2\pi \ri {\boldsymbol \xi} }-{1\over (2\pi \ri)}{\boldsymbol T}\right)^n\Theta =\left( \prod_{I=1}^{n} \left(\partial_{2\pi \ri { \xi_{p_I}} }-{1\over (2\pi \ri)}{ T_{p_I}}\right)\right)\Theta =
\Theta^{(n)}.\ee
The modular properties of $\Theta^{(n)}$ have been worked out in \cite{em} and they can be schematically represented as
\be\label{tt}\ba {\Theta^{(n)}\over \Theta} \to (C\widehat \tau+D)^n\sum_{j=0}^{[n/2]}(2j-1)!!\binom{n}{2j}\left(-{ {\mathcal K} \over 2 \pi \ri  }\right)^{j} {\Theta^{(n-2j)}\over \Theta},\ea \ee
where $ {\mathcal K}$ is given in \eqref{kappahat}.
Notice that when we refer to modular invariance we always mean up to a change of characteristic in the theta function \eqref{thetadef} as in \cite{em}.
In the following we will omit the constant map contribution $A(\hbar)$ appearing in the spectral determinant \eqref{Jtot} since it clearly does not affect the modular properties. To be more concrete we will do the analysis in the case of a mirror curve with genus one.

\subsection{First order}
Let us look at the first order in the $\hbar-2\pi$ expansion. 
After some computations we find
\be \ba \partial_{\hbar}J(\mu+2\pi \ri n, \hbar)\Big|_{\hbar=2\pi}=&\left(n-{\widehat T \over 2\pi \ri}\right) \left( \ri \partial_T \widehat F_1  + \ri \partial_T \widehat F_1^{\rm  NS}\right)+\frac{1}{6} \ri \partial^3_T \widehat{F}_0 \left(n-{\widehat T \over 2\pi \ri}\right)^3. \ea\ee
It follows that
\be \label{1o} \ba {\rd \over \rd \hbar} \Xi(\kappa, \hbar)\Big|_{\hbar=2\pi}
&=  \re^{N^2 {\widehat F_0} +\widehat F_1-\widehat F_1^{\rm NS}} \Big(  {\ri\over 6}\partial_T ^3 {\widehat F_0} \Theta^{(3)} + \ri   \left( \partial_T \widehat F_1+\partial_T \widehat  F_1^{\rm NS}\right)\Theta^{(1)}\Big),
\ea \ee
where  we used $\partial_{\hbar} T \mid_{\hbar=2\pi}=0$.
By using the modular properties of the unrefined free energy \cite{abk,em}, together with \eqref{tt} we have 
\be \ba \partial_T ^3 {\widehat F_0}  \to& {1\over (C\widehat \tau+D)^3} \partial_T ^3 {\widehat F_0},\\
 \partial_T \widehat F_1 \to &(C\widehat \tau+D)^{-1} \left( \partial_T \widehat F_1+{1\over 2}{ {\mathcal K}\over 2 \pi \ri} \partial_T ^3 {\widehat F_0}\right),\\
{\Theta\over \Theta}^{(1)} \to &(C \widehat \tau+D) {\Theta\over \Theta}^{(1)},\\
{\Theta\over \Theta}^{(3)} \to &(C \widehat\tau+D)^3 \left( {\Theta\over \Theta}^{(3)}-{3}{ {\mathcal K} \over 2 \pi \ri }{\Theta\over \Theta}^{(1)} \right).  \ea \ee
 From the invariance of  $F_1^{\rm NS}$  it follows that
\be \partial_T \widehat F_1^{\rm NS}\to\left( C \widehat \tau +D\right)^{-1} \partial_T \widehat F_1^{\rm NS}.\ee
Therefore \eqref{1o} is manifestly modular invariant as expected.
\subsection{Second order}
At the second order we have
\be \ba {d^2 \over d \hbar^2}  \Xi(\kappa, \hbar)\Big|_{\hbar=2\pi}=\sum_n& \left( \partial_{\hbar}^2J(\mu+2\pi \ri n, \hbar) +\left(\partial_{\hbar}J(\mu+2\pi \ri n, \hbar)\right)^2+\partial_{\hbar}^2T \partial_T J(\mu+2\pi \ri n, \hbar) \right)\Big|_{\hbar=2\pi}\\
& \times \re^{J(\mu+2\pi \ri n, 2\pi)} \ea \ee
One can show from \eqref{Jtot} that
\be \ba \partial_{\hbar}^2J(\mu+2\pi \ri n, \hbar)\big|_{\hbar=2\pi}= &-\frac{1}{12}\left(n-{\widehat T \over 2\pi \ri}\right)^4 \partial_T^4 \widehat F_0-\frac{\ri \left(n-{\widehat T \over 2\pi \ri}\right)^3 \partial_T^3 \widehat F_0}{6 \pi }\\
&-\left(n-{\widehat T \over 2\pi \ri}\right)^2 \partial_T^2 \widehat F_1 -{\ri\over\pi}\left(n-{\widehat T \over 2\pi \ri}\right)\partial_T \widehat F_1\\
&-2\widehat F_2-{6}\widehat F_2^{\rm NS}.\ea\ee
Similarly 
\be \partial_{\hbar}^2 T \partial_T J (\mu+2\pi \ri n,\hbar)\big|_{\hbar=2\pi}=2\widehat T^{(1)}\left( \partial_T \widehat F_1- \partial_T \widehat F_1^{\rm NS}+{1\over 2 }\partial_T^3 \widehat F_0 (n-{\widehat T \over 2 \pi \ri})^2 \right), \ee
where we use that \be \ba  &\partial_{\hbar}^2 T \mid_{\hbar=2\pi} (z)=\partial_{\hbar}^2 T \mid_{\hbar=0} ((-1)^{B}z)=2\widehat T^{(1)},\\
\ea \ee
and $B$ is the $B$ field. 
It follows that
\be \label{sec} \ba {1 \over 2}{d^2 \over d \hbar^2}  \Xi(\kappa, \hbar)\Big|_{\hbar=2\pi}=&  \re^{N^2 {\widehat F_0} +\widehat F_1-\widehat F_1^{\rm NS}} \Big(\left(- \widehat F_2 -3\widehat F_2^{\rm NS} \right)\Theta
 -{1\over 72}\left(\partial_T ^3 {\widehat F_0} \right)^2\Theta^{(6)}- {1\over 24}\partial_T ^4 {\widehat F_0} \Theta^{(4)}
\\&- {1\over 6}\partial_T ^3 {\widehat F_0}\left(\partial_T \widehat F_1+\partial_T  \widehat F_1^{\rm  NS}\right) \Theta^{(4)}-{\ri \over 2 \pi } \left( {1\over 6}\partial_T ^3 {\widehat F_0} \Theta^{(3)}+\partial_T {\widehat F_0} \Theta^{(1)}\right)\\&-{1\over 2} \left((\partial_T \widehat F_1+\partial_T \widehat F_1^{\rm  NS})^2+\partial_T ^2\widehat F_1\right)\Theta^{(2)}
\\
&+\widehat T^{(1)}\left( \partial_T \widehat F_1 \Theta- \partial_T \widehat F_1^{\rm NS} \Theta+{1\over 2 }\partial_T ^3 {\widehat F_0}  \Theta^{(2)} \right)\Big).
\ea \ee
When the NS quantities are set to zero, modular invariance of \eqref{sec} is straightforward from \cite{em}.
Therefore it is enough to show that the following term 
is  invariant under \eqref{cytra}:
\be \label{mttd}\ba  \re^{N^2 {\widehat F_0} +\widehat F_1-\widehat F_1^{\rm NS}} &\Big(- 3  \widehat F_2^{\rm NS}  \Theta  - {1\over 6}\partial_T ^3 {\widehat F_0}\partial_T \widehat F_1^{\rm NS}\ \Theta^{(4)} -{1\over 2} \left(\left(\partial_T \widehat F_1 ^{\rm NS}\right)^2+2\partial_T \widehat F_1 ^{\rm NS}\partial_T \widehat F_1 \right)\Theta^{(2)}\\
&+\widehat T ^{(1)}\left( \partial_T  \widehat F_1 \Theta^{(0)}- \partial_T  \widehat F_1^{\rm NS} \Theta^{(0)}+{1\over 2 }\partial_T ^3 {\widehat F_0}  \Theta^{(2)} \right)\Big). \ea \ee
By using the transformation properties of the standard free energy \cite{abk,em,emo} together with \eqref{qpt} and \eqref{tt} we find
\be \label{ft} {1\over 6}\partial_T^3 {\widehat F_0}\left(\partial_T \widehat F_1^{\rm NS}\right) \Theta^{(4)}/\Theta \to\partial_T ^3 {\widehat F_0}\left(\partial_T \widehat F_1^{\rm NS}\right)\left( {1\over 6}{ \Theta \over \Theta}^{(4)}- { {\mathcal K}\over 2\pi \ri} {\Theta\over \Theta}^{(2)}+{1\over 2}\left({ {\mathcal K}\over 2 \pi \ri}\right)^2   \right).\ee
Similarly 
\be \label{st} \ba \partial_T \widehat F_1 ^{\rm NS}\partial_T \widehat F_1 \Theta^{(2)} /\Theta \to & \partial_T \widehat F_1 ^{\rm NS}\partial_T \widehat F_1  \left({\Theta\over \Theta}^{(2)}-{  {\mathcal K} \over 2\pi \ri }\right) \\
&+{1\over 2} {  {\mathcal K}\over 2 \pi \ri}  \partial_T \widehat F_1 ^{\rm NS}\partial_T^3 \widehat F_0  \left({\Theta \over \Theta}^{(2)}-{  {\mathcal K} \over 2\pi \ri }\right).  \ea\ee 
Moreover 
\be\label{tttt}- {1\over 2}\left(\partial_T \widehat F_1^{\rm NS}\right)^2 \Theta^{(2)}/ \Theta   \to - \left(\partial_T \widehat F_1^{\rm NS}\right)^2 \left( {1\over 2} {\Theta\over \Theta}^{(2)}  -{1\over 2}{ {\mathcal K}\over 2\pi \ri}  \right), \ee
and
\be \label{fht} 3  \widehat F_2^{\rm NS}  \to  3  \widehat F_2^{\rm NS}  + 3 { {\mathcal K} \over 2}\left({1\over (2\pi \ri)^{1/2}}\partial_T \widehat F_1^{\rm NS}\right)^2 . \ee
By summing up \eqref{ft},\eqref{st},\eqref{tttt} and \eqref{fht} we obtain modular invariance up to an additional term given by
\be \label{plus}{ {\mathcal K}\over 2\pi \ri} \partial_T \widehat F_1 ^{\rm NS}\left(-  \partial_T \widehat F_1 ^{\rm NS}+\partial_T \widehat F_1\right)+{1\over 2} {  {\mathcal K}\over 2\pi \ri}  \partial_T \widehat F_1 ^{\rm NS}\partial_T^3 \widehat F_0  {\Theta \over \Theta}^{(2)}. \ee
For the last term in \eqref{mttd} we have
\be \label{ep1t}\ba \widehat T^{(1)}\left( \partial_T  \widehat F_1 + \partial_T  \widehat F_1^{\rm NS}+{1\over 2 }\partial_T ^3 {\widehat F_0}  {\Theta \over \Theta}^{(2)} \right) \to &\widehat T^{(1)}\left( \partial_T  \widehat F_1 - \partial_T  \widehat F_1^{\rm NS} +{1\over 2 }\partial_T ^3 {\widehat F_0} { \Theta\over \Theta}^{(2)} \right)\\
&-\partial_T \widehat F_1^{\rm NS} { {\mathcal K}\over 2 \pi \ri}  \left( {1\over 2} \partial_T^3\widehat F_0 {\Theta\over \Theta}^{(2)} +\partial_T \widehat F_1-\partial_T \widehat F_1^{\rm NS}\right).
\ea\ee
In particular the second term in \eqref{ep1t} precisely cancels the one in \eqref{plus}.
Hence modular invariance is manifest also at the second order.

\subsection{The general mechanism} \label{qftg}

The general structure of the $n^{\rm th}>0$ order term in the expansion \eqref{series} is the following. There is a part which involves only standard topological strings 
\be \Xi(\kappa, 2\pi) \prod_i^k \partial ^{n_i}_T \widehat F_{g_i} \Theta^{(n_1+\cdots+n_k)}/ \Theta, \qquad g_i \leq n.\ee
By using the combinatorial rules of \cite{em} it follows that this part  is modular invariant.
 However we also have two other contributions of the form ($g_i, g_j < n$)
\be \label{NN1}  \Xi(\kappa, 2\pi) \prod_{i}^k \partial ^{n_i}_T\widehat F_{g_{i}}\prod_{j}^\ell  \partial ^{m_j}_T\widehat F_{g_j} ^{\rm NS}\Theta^{(n_1+m_1+\cdots+m_\ell+n_k)}/ \Theta , \ee 
 \be\label{NN2} \Xi(\kappa, 2\pi)\prod_p^s \widehat T^{(r_p)} \prod_{i}^k \partial ^{n_i}_T\widehat F_{g_{i}}\prod_{j}^\ell  \partial^{m_j}_T\widehat F_{g_j} ^{\rm NS} \Theta^{(n_1+m_1\cdots+m_\ell+n_k-s)}/ \Theta.  \ee 
Let us look at the first term \eqref{NN1}. Since the number of derivatives acting on $\Theta$ is the same as acting on the free energy, when we apply a symplectic transformation we
 recover the same term plus an extra factor. The same happens also for the second term \eqref{NN2}. Indeed $\widehat T^{(k)}$ get a multiplicative factor of  $(C \widehat \tau +D)$  under modular transformations as explained in section \ref{sepsi}. Notice that  \eqref{NN2} appears when we consider the  $\hbar$ dependence of $ T$. 

  None of the  terms \eqref{NN1} and \eqref{NN2} is modular  invariant: they both transform into itself  plus an additional factor. At each order in the  $\hbar-2\pi$ expansion, the extra factor coming from \eqref{NN1} cancels the one coming from \eqref{NN2}. Although we don't have a general proof of this cancellation, 
   we have checked it by an explicit computation up to order 4. The details of the computations at order 3 are given in Appendix 
   \ref{apo3}.
  
  \section{Testing the topological string/quantum mechanics duality}

  By now several tests of the conjecture  \cite{ghm} have been done. In the analytic tests an important roles is played by the Fermionic spectral traces
  \be \label{spta} Z({\bf N}, \hbar)\ee
  appearing in the small $\kappa$ expansion of the spectral determinant \eqref{xitoN}. These can be computed analytically on both side of the conjecture.
 In the topological string side this is done by computing explicitly the small $\kappa$ expansion  of \eqref{sdtot} while
 in the operator theory side one gets an analytic expression for  \eqref{spta} by computing the traces of the operators \eqref{rhoop} as we will explain below.
   
  In this section we are interested in the derivative of the Fermionic spectral traces
    \be \label{spta2} {\rd \over \rd \hbar}Z({\bf N}, \hbar).\ee

 Let us consider the simplest example, namely the local $\IP^2$ geometry.
The small $\kappa$ expansion \eqref{xitoN} of the spectral determinant at $\hbar=2\pi$ was computed in \cite{ghm} and reads
\be   \Xi(\kappa, 2 \pi) =1+{1\over 9} \kappa +\left( {1\over 12 \pi \sqrt{3}}-{1\over 81}\right)\kappa^2+\left({5 \over 2187} - {1\over 72 \pi^2} -{1 \over 216 \pi \sqrt{3}}\right)\kappa^3+\mathcal{O}(\kappa^4).\ee
Similarly by using \eqref{sdtot}, \eqref{1o}  we have 
\be  \label{dxp2}\ba {\rd \over \rd \hbar} \Xi(\kappa, \hbar)\mid_{\hbar=2\pi}
=  \Xi(\kappa, 2 \pi)& \Big( A'(2\pi)-  {9 \ri \over 2 }\partial_t ^3 {\widehat F_0} {\left( {1\over 2 \pi \ri}\partial_{\xi}-{\ri \over 6 \pi }t(\kappa, 2\pi) \right)^3 \vartheta_2(\tau_{\rm or}(\kappa), \xi(\kappa)-1/4) \over \vartheta_2(\tau_{\rm or}(\kappa), \xi(\kappa)-1/4)} \\
&- {3\ri }   \left( \partial_t \widehat F_1+\partial_t \widehat  F_1^{\rm NS}\right){\left( {1\over 2 \pi \ri}\partial_{\xi}- {\ri \over 6 \pi }t(\kappa,2\pi)\right)\vartheta_2(\tau_{\rm or}(\kappa),  \xi(\kappa)-1/4)\over \vartheta_2(\tau_{\rm or}(\kappa)), \xi(\kappa)-1/4)}\Big),\ea \ee
where we denote by  $\vartheta_2(\tau, v) $
the Jacobi theta function of characteristic two  
\be \vartheta_2(\tau, v)=\sum_{n \in \IZ}\re^{\ri \pi (n+1/2)^2 \tau+2\pi \ri (n+1/2)v} \ee
and we use
\be \tau_{\rm or}(\kappa)=-{9 \ri \over 2 \pi}\partial^2_t \widehat F_0(\kappa)-{1\over 2}, \qquad \xi(\kappa)=\frac{3}{4 \pi ^2}\left( -t(2\pi) \partial^2_t \widehat F_0(\kappa)+\partial_t \widehat F_0(\kappa) \right).\ee
To compute the small $\kappa$ expansion of \eqref{dxp2} one has to analytically continue the free energy and the K\"ahler parameter to the orbifold point. By using the expression of appendix \ref{apor}, \ref{applr}   for $A(\hbar)$, $\partial_t ^3 {\widehat F_0}$, $ \widehat F_1$ and $ \widehat F_1^{\rm NS}$ we find 
\be \label{std33}\ba {\rd \over \rd \hbar} \Xi(\kappa, \hbar)\mid_{\hbar=2\pi}=& -\left(\frac{1}{18 \pi }+\frac{1}{54 \sqrt{3}}\right)\kappa+  \frac{\pi  \left(99-16 \sqrt{3} \pi \right)-81 \sqrt{3}}{5832 \pi ^2}\kappa^2
\\ & + \frac{\pi  \left(5 \pi  \left(9-8 \sqrt{3} \pi \right)+162 \sqrt{3}\right)+729}{104976 \pi ^3}\kappa^3 + \mathcal{O}(\kappa^4). \ea\ee
To derive this result one has to use non trivial identities involving derivatives of the theta function evaluated at some special points. For instance we found that 
\be \partial_v^3 \vartheta_2 \left(\re^{2\pi \ri/3}, -{1\over 6}\right)=-\frac{8 (-1)^{7/8} \pi ^2 A'(2\pi) \Gamma \left(\frac{1}{3}\right)^{9/2}}{3^{19/24} \Gamma \left(\frac{2}{3}\right)^6}-2 \sqrt{3} \pi  \partial_v \vartheta_2 \left(\re^{2\pi \ri/3}, -{1\over 6}\right),\ee
where  \be A'(2\pi)= -\frac{{\rm Im}\left(\text{Li}_2\left(\frac{\re^{-\frac{\ri \pi}{6} }}{\sqrt{3}}\right)\right)}{4 \pi ^2}+\frac{2}{27 \sqrt{3}}+\frac{\log (3)}{48 \pi }-\frac{\psi ^{(1)}\left(\frac{1}{3}\right)}{12 \sqrt{3} \pi ^2}\ee
and we denote by $\psi ^{(1)}$ the polygamma function of order one (see appendix \ref{apor}).
To our knowledge  this identity is not known in the literature however one can easily check it numerically with arbitrarily high precision.

Therefore, on the  string side of the duality \cite{ghm} we have
\be \label{std3}\ba  &{\rd \over \rd \hbar} Z(1, \hbar) \big|_{\hbar=2 \pi}=-\frac{1}{18 \pi }-\frac{1}{54 \sqrt{3}}, \\
& {\rd \over \rd \hbar} Z(2, \hbar) \mid_{\hbar=2 \pi}=\frac{\pi  \left(99-16 \sqrt{3} \pi \right)-81 \sqrt{3}}{5832 \pi ^2}. \ea  \ee
Alternatively, these results can  be derived by using the Airy function method, namely
\be \label{airyz} {\rd \over \rd \hbar} Z(N, \hbar)={1\over 2 \pi \ri} \int_{\mathcal{C}}\rd \mu  \re^{J(\mu, \hbar)- \mu N} {\rd   \over \rd \hbar}\left( J(\mu,  \hbar) \right), \ee
where $\mathcal{C}$ is the standard Airy contour. 
It is easy to check that a numerical evaluation of the integral \eqref{airyz}  with the method of  \cite{hmo2} leads to  \eqref{std33}.

We are now going to reproduce \eqref{std3} in the operator side of the duality. 
The kernel of  \eqref{thop2}  was computed in  \cite{kama} and reads
 \be\label{kernelp2} \rho_{\IP^2}(p,p')=\frac{\overline{\mypsi{a}{a}(p)}\mypsi{a}{a}(p')}{2\mathsf{b}\cosh\left(\pi\frac{p-p'+\im (a)}{\mathsf{b}}\right)},
\quad a={b\over 6}, \quad \hbar={2\pi \over 3}b^2, \ee
 where \cite{ak}
 \be \mypsi{a}{c}(x)= \frac{\re^{2\pi ax}}{\fad(x-\im(a+c))} \ee
 and  $\fad$ is  the Faddeev's quantum dilogarithm \cite{faddeev, fk} whose integral representation is 
\be
\fad(x)=\exp \left( \int_{\mathbb{R}+\im\epsilon}
\frac{\re^{-2\im xz}}{4\sinh(z\mathsf{b})\sinh(z\mathsf{b}^{-1})}
{\operatorname{d}\!z  \over z} \right), \quad |\mathrm{Im} z| < |\mathrm{Re} \, { {b+b^{-1} \over 2}}|.
\ee
 The spectral traces for the operator $\rho_{\IP^2}$ are defined by
 \be \label{fstst}\ba Z_{\ell} (\hbar)=\sum_{n\geq 0} \re^{-\ell E_n}= \Tr \widehat \rho_{\IP^2}^{\ell}= \int \rd ^{\ell }x \prod_{i=1}^\ell   \rho_{\IP^2}(x_i,x_{i+1}),  \quad x_{\ell+1}=x_1
\ea\ee
where $\re^{-E_n}$ denotes the spectrum of the operator \eqref{thop2}.
It was shown in  \cite{kama}  that 
\be \label{km}\ba& Z_1(\hbar)= {1\over \mathsf{b}}\left|\fad\left( \frac{\ri \left(b^2+3\right)}{6 b}\right)\right|^3,  \\
& Z_2(\hbar)= \frac{\left|\fad\left((\ri \left(b^2-3\right))/{6 b}\right)\right|^2}{\sqrt{3}b }\int_{\mathbb{R}}\frac{\sinh(\pi s b/3)}{\sinh(\pi{b}s)}\left|\mypsi{A}{A}(x)\right|^4 , \quad A=\frac{b^2+3}{12 b}.
\ea\ee
It follows that
\be\label{sp1} \ba {\rd \over \rd \hbar}Z_1(\hbar) \mid_{\hbar=2 \pi}=&-{1\over 4 \pi} \left({1\over \sqrt{3}}+\int_{\IR} \rd t \frac{4 \cosh ^2\left(\frac{t}{\sqrt{3}}\right)}{\left(2 \cosh \left(\frac{2 t}{\sqrt{3}}\right)+1\right)^2} \right) \Big|  \fad\left(\frac{\ri \left(b^2+3\right)}{6 b} \right)\Big|^3 \Big|_{b=\sqrt{3}}\\
=&-\frac{1}{18 \pi }-\frac{1}{54 \sqrt{3}}\ea\ee
where we used 
\be
\ba 
\fad\left( \frac{\ri \left(b^2+3\right)}{6 b}\right) \Big|_{b=\sqrt{3}}=\frac{\exp \left(- \ri \pi /36  \right)}{\sqrt{3}}.
\ea
 \ee
 Similarly
\be \label{sp2}\ba 
{\rd \over \rd \hbar}Z_2(\hbar) \Big|_{\hbar=2 \pi}&={\sqrt{3}\over 4 \pi} Z_2(2\pi) \left(\int_{\IR}\rd t\frac{\text{csch}^2\left(\frac{t}{\sqrt{3}}\right)}{6 \cosh \left(\frac{2 t}{\sqrt{3}}\right)+3}-{1\over \sqrt{3}}\right) - {1 \over  3 \sqrt{3} } \int_{\IR} \rd t \frac{t \sinh \left(\frac{2 \pi  t}{\sqrt{3}}\right)}{ \left(2 \cosh \left(\frac{2 \pi  t}{\sqrt{3}}\right)+1\right)^4}\\
&-{4\over 3 \sqrt{3} \pi}\int_\IR \rd s \int_{\IR } \rd t \frac{e^{2 \ri s t}}{  \left(2 \cosh \left(\frac{2 \pi  s}{\sqrt{3}}\right)+1\right)^3} \frac{\cos ^2\left(\frac{1}{6} \left(\pi +2 \sqrt{3} \ri t\right)\right)}{\left(1+2 \cos \left(\frac{1}{3} \left(\pi +2 \sqrt{3} \ri t\right)\right)\right)^2}\\
&=\frac{\pi  \left(4 \sqrt{3} \pi -45\right)+27 \sqrt{3}}{972 \pi ^2},\ea \ee
where we used the properties of the quantum dilogarithm to write  (see for instance \cite{garou-kas},\cite{ak}) 
\be \Big| \mypsi{1/(2\sqrt{3})}{1/(2\sqrt{3})}(s)\Big|^2={\sinh(\pi s / \sqrt{3})\over \sinh(\pi s \sqrt{3})}.\ee
Moreover
\be \int_{\IR } \rd t \frac{\cosh ^2\left(\frac{t}{\sqrt{3}}\right)}{\left(2 \cosh \left(\frac{2 t}{\sqrt{3}}\right)+1\right)^2} \int_\IR \rd s \frac{e^{2 \ri s t}}{  \left(2 \cosh \left(\frac{2 \pi  s}{\sqrt{3}}\right)+1\right)^3} =\frac{33 \sqrt{3}-16 \pi }{1296}.\ee
By using the relation between  the spectral traces \eqref{fstst} and the Fermionic spectral traces \eqref{spta} 
\be \ba Z(1, \hbar)&=Z_1(\hbar), \\  Z(2, \hbar)&={1\over 2} (Z_1^2(\hbar) - Z_2(\hbar)), \\
 \ea\ee
 it is easy to check that \eqref{sp1}, \eqref{sp2}  reproduce \eqref{std3} as expected from the topological strings/spectral theory duality.

\section{Conclusions}

We studied the behavior of the  Nekrasov--Shatashvili  free energies under a change of symplectic frame and we found that it can be organized by means of  simple diagrammatic rules.  
We used these results to investigate the symplectic properties of the spectral determinants of operators associated to mirror curves.  In turn, these  can be expressed in term of a quantum theta 
\be\label{qtfin} \Theta_X({\boldsymbol \kappa}, \hbar)\ee
which,  at least as an asymptotic expansion,  is manifestly well defined  and  has a clear non--perturbative meaning as a spectral determinant \cite{ghm,cgm2}. 
 In particular when $\hbar=2\pi$, this quantum theta function becomes a conventional theta function whose modular properties are well known.  
 By expanding around this special point we found that,   order by  order  the $\hbar-2\pi$ expansion, the corresponding spectral determinants are invariant under a change of symplectic frame. 
In view of these results,  we provided a new test the conjecture  \cite{ghm}.%

In this paper we performed a further step in understanding the general properties of the quantum theta function away from the large radius region of the moduli space. However to have a complete understanding of this object it would be interesting to find a closed form  expression for \eqref{qtfin}  at any point in the moduli space and for all values of $\hbar$. This would require a Goparkumar--Vafa like resummation away from the large radius point which, at present, it is not known.

Moreover we observed that the Nekrasov--Shatashvili partition function  behaves as a wave function and it is  well known \cite{bkmp,mmopen} that in the unrefined case this kind of behavior   is closely related to the  topological recursion of Eynard and Orantin  \cite{eo}. Therefore,
given the above diagrammatic rules, one would expect that it exists a simple  "refined" topological recursion also for the NS free energies
 compatible with \eqref{fid}, \eqref{sid}, \eqref{tid}.  
We hope to report on this in the near future.
 
\section*{Acknowledgements}
I am grateful to Marcos Mari\~no  for the many and useful discussions,  for suggesting me to investigate the behavior of the spectral determinant around the special value of $\hbar=2\pi$ and for a detailed reading of the draft.
I would like to thank Teresa Bautista,  Atish Dabholkar, Antonio Sciarappa, Alessandro Tanzini,  and Szabolcs Zakany for useful discussions and comments on the draft.

\appendix
\section{The local $\IP^2$ geometry} \label{app2}
In this section we collect some results on the local $\IP^2$ geometry.
\subsection{The large radius point} \label{applr}
The quantum A period  for the local $\IP^2$ geometry has been computed in \cite{acdkv} and it reads 
\be \ba t(z, \hbar)=& -{3 \over (2 \pi \ri)^{1/2}}\epsilon(z,\hbar)=-3 T(z,\hbar)=-\log (z) +6 z \cos \left(\frac{\hbar }{2}\right)- 3 z^2 (7 \cos (\hbar )+2 \cos (2 \hbar )+6)
\\ &+2 z^3 \left(144 \cos \left(\frac{\hbar }{2}\right)+88 \cos \left(\frac{3 \hbar }{2}\right)+36 \cos \left(\frac{5 \hbar }{2}\right)+9 \cos \left(\frac{7 \hbar }{2}\right)+3 \cos \left(\frac{9 \hbar }{2}\right)\right)+O\left(z^4\right). \ea \ee
Hence \be  t(z, \hbar)= \sum_{n \geq 0}\hbar^{2n}t^{(n)},\ee
where
\be \label{epeg}\ba 
t^{(0)}=& -{3 \over (2 \pi \ri)^{1/2}}\epsilon^{(0)}= -\log (z)+6 z -45 z^2+560 z^3+O\left(z^4\right),\\
t^{(1)}=& -{3 \over (2 \pi \ri)^{1/2}}\epsilon^{(1)}=- \frac{3 z}{4}-630 z^3+{45\over 2} z^2+O\left(z^4\right),\\
t^{(2)}=& -{3 \over (2 \pi \ri)^{1/2}}\epsilon^{(2)}={1 \over 64}z - {39\over 8} z^2 + {2961\over 8} z^3+O\left(z^4\right), \\
&\qquad \qquad\qquad\qquad \quad  \vdots
\ea\ee
The NS free energies of the local $\IP^2$ geometry  be computed by using the refined topological vertex \cite{akmv2} or the refined holomorphic anomaly \cite{hk,hkpk, huangNS}. We have \footnote{ These are given at $\hbar$=0.}
\be \label{idfr} \ba F_0(t)=&-{1\over 18}t^3-\sum_{j_L, j_R} \sum_{w, {\bf d} } N^{{\bf d}}_{j_L, j_R}  \frac{(2 j_L+1) (2 j_R+1) (-1)^{2 j_L+2 j_R} \re^{-d w t}}{w^3}\\
=&\frac{\log ^3(z)}{18}+z \left(-\log ^2(z)-3\right)+z^2 \left(\frac{15 \log
   ^2(z)}{2}+6 \log (z)+\frac{189}{8}\right)+O\left(z^3\right)\ea\ee
\be F_1^{\rm NS}(t)={1\over 24}t+\sum_{j_L, j_R} \sum_{w, {\bf d} } N^{{\bf d}}_{j_L, j_R} \frac{ m_L  m_R \left( m_L^2+ m_R^2-3\right) \re^{-d wt}}{24 w}= \frac{1}{24} \log \left(\frac{27 z+1}{z}\right)\ee
\be \ba F_2^{\rm NS}(t)=&- \sum_{j_L, j_R} \sum_{w, {\bf d} } N^{{\bf d}}_{j_L, j_R}\frac{  m_L  m_R w (-1)^{ m_L+ m_R}\re^{-d wt})}{5760} \left(3  m_L^4+10  m_L^2 \left( m_R^2-3\right)+3  m_R^4-30  m_R^2+51\right)\\
&=-\frac{29 z}{640}+\frac{561 z^2}{160}-\frac{20091 z^3}{128}+ \frac{1898037 z^4}{320}+\mathcal{O}(z^5)\ea\ee
\be\ba  F_3^{\rm NS}(t)=- &\sum_{j_L, j_R} \sum_{w, {\bf d} } N^{{\bf d}}_{j_L, j_R} \frac{ m_L  m_R w^3 (-1)^{ m_L+ m_R+1} \re^{-d w t}}{967680}\\
& \left(3 \left( m_L^6+7  m_L^4 \left( m_R^2-3\right)+7  m_L^2 \left( m_R^4-10  m_R^2+17\right)+ m_R^4 \left( m_R^2-21\right)\right) +357  m_R^2-457 \right)\\
=&-\frac{101329821 z^4}{17920}+\frac{2678163 z^3}{35840}-\frac{1449 z^2}{2560}+\frac{137 z}{322560}+\mathcal{O}(z^5),\ea \ee
where we denote $t= t^{(0)}$  and we use the BPS $N^{{\bf d}}_{j_L, j_R} $ invariants  listed in \cite{ckk}.

\subsection{An explicit computation}\label{soid}
In this section we use the computation of \cite{huangNS} to check explicitly  the modular transformations  \eqref{ept}.
We consider the example of local $\IP^2$. In this case there is one single complex moduli $z=\kappa^{-3}$ and the operators in  \eqref{huang} read
\be \label{epi}\ba & \epsilon^{(i)}=\mathsf{D}_1^{(i)}\epsilon^{(0)}+\mathsf{D}_2^{(i)}\epsilon^{(0)},\\
\ea\ee
where
\be \ba \mathsf{D}_1^{(1)}= 0, &\qquad \mathsf{D}_2^{(1)}=-{1\over 8}(z\partial_z)^2 ,\\
\mathsf{D}_1^{(2)}= {2z(999z-5)\over 640 (1+27 z)^2}z\partial_z, &\qquad \mathsf{D}_2^{(2)}= {3z(2691z-29)\over 640 (1+27 z)^2}(z\partial_z)^2,\ea \ee
\be \ba
\mathsf{D}_1^{(3)}= &\frac{z^2 (7-3 z (459 z (18657 z-2281)+9677)) }{53760 (27 z+1)^4} z\partial_z,  \\
\quad \mathsf{D}_2^{(3)}=& \frac{z (137-243 z (9 z (80793 z-12565)+1495))}{107520 (27 z+1)^4}(z\partial_z)^2, \\
&\qquad \qquad\qquad\qquad \quad  \vdots
\ea \ee
For the local $\IP^2$ geometry  the $\epsilon^{(i)}$ are given in \eqref{epeg}.
It follows that
\be  \epsilon^{(1)}= \mathsf{D}_2^{(1)}\epsilon^{(0)}. \ee
By using the transformation of the classical periods \eqref{clp} together with the relation found in 
\cite{huangNS}
\be  ( z \partial_z)^2 \partial F_0 = \tau  ( z \partial_z )^2\epsilon^{(0)}+ \partial^3 F_0 ( z \partial_z \epsilon^{(0)})^2 ,\ee
we get
 \be \label{idd1} \ba \epsilon^{(1)}\quad \to &\quad \left(C\tau +D\right)\epsilon^{(1)}-{1\over 8}\partial^3F_0 ( z \partial_z \epsilon^{(0)})^2.\ea \ee
 However in  \eqref{ept} we have that
 \be  \label{idd2} \epsilon^{(1)} \quad \to  \quad C \left(\partial  F_{1}^{\rm NS}+ \epsilon^{(1)}\tau\right) +D\epsilon^{(1)} .\\
\ee
For the  local $\IP^2$ geometry it is easy to check that
\be\label{fid} \partial^3_\epsilon F_0(z\partial_z \epsilon^{(0)})^2=-{1\over 8}\partial_\epsilon F_1^{\rm NS}. \ee
 where the  expressions for $F_0, F_1^{\rm NS}, \epsilon$ are given in  appendix \ref{app2}. Therefore the transformation  \eqref{idd2} derived by using diagrammatic rules  matches precisely the one obtained by explicit computation.

 Likewise an explicit computation along the way of \cite{huangNS} leads to
 \be \label{idd22}\epsilon^{(2)}\quad \to \quad \left(C\tau +D\right)\epsilon^{(2)}+ C \left( \frac{3 z (2619 z-29)}{640 (27 z+1)^2} \partial^3_\epsilon F_0(z\partial_z \epsilon^{(0)})^2\right) \ee
The diagrammatic rules \eqref{ept} instead leads to
\be \label{idd21}\epsilon^{(2)} \to C \left(\partial  F_{2} ^{\rm NS}+\epsilon^{(1)}\partial ^2 F_{1} ^{\rm NS}+{1\over 2} (\epsilon^{(1)} )^2\partial ^3 F_{0} + \epsilon^{(2)}\tau \right) +D\epsilon^{(2)}\ee
Moreover by using \eqref{idfr}  it is easy to check that the following identity holds \be\label{sid} \ba \frac{3 z (2619 z-29)}{640 (27 z+1)^2} \partial^3_\epsilon F_0(z\partial_z \epsilon^{(0)})^2= &\partial  F_{2} ^{\rm NS}+\epsilon^{(1)}\partial ^2 F_{1} ^{\rm NS}+{1\over 2} (\epsilon^{(1)} )^2\partial ^3 F_{0}.  \ea  \ee
This leads to a complete agreement between  \eqref{idd22} and \eqref{idd21}.
Similarly, by using
\be \label{tid} \ba &-\frac{z \left(\left(176694291 z^3-27479655 z^2\right)+(363285 z-137)\right)}{107520 (27 z+1)^4}\partial^3_\epsilon F_0(z\partial_z \epsilon^{(0)})^2=\\
&\frac{1}{6}(\epsilon^{(1)})^3 \partial^4 F_0+\epsilon^{(1)}\epsilon^{(2)} \partial^3 F_0 \frac{1}{2}(\epsilon^{(1)})^2 \partial^3 F_1^{\rm NS}+\epsilon^{(2)}\ \partial^2 F_1^{\rm NS}+\epsilon^{(1)}\partial^2F_2^{\rm NS}+\partial F_3^{\rm NS},\ea  \ee
we can check by a direct computation  the modular transformation of $\epsilon^{(3)}$ as given in \eqref{ept}.

\subsection{Analytic continuation to the orbifold point} \label{apor}
To compute the Fermionic spectral traces \eqref{spta} from the spectral determinant \eqref{xitoN}  it is useful to analytically  continue the free energy and the K\"ahler parameter to the orbifold region where 
\be \kappa \ll 1.\ee
For the local $\IP^2$ geometry this was be done for instance in \cite{abk}. One finds
\be\label{tor} t(z,0)=\frac{\Gamma \left(\frac{2}{3}\right) \, _3F_2\left(\frac{2}{3},\frac{2}{3},\frac{2}{3};\frac{4}{3},\frac{5}{3};-\frac{1}{27 z}\right)}{2 z^{2/3} \Gamma \left(\frac{1}{3}\right)^2}-\frac{\Gamma \left(\frac{1}{3}\right) \, _3F_2\left(\frac{1}{3},\frac{1}{3},\frac{1}{3};\frac{2}{3},\frac{4}{3};-\frac{1}{27 z}\right)}{\sqrt[3]{z} \Gamma \left(\frac{2}{3}\right)^2} ,\ee
\be\label{F0ro} \ba  \partial_t F_0(z)&=\frac{1}{9} \pi  \left(\sqrt{3} \left(\frac{\Gamma \left(\frac{2}{3}\right) \, _3F_2\left(\frac{2}{3},\frac{2}{3},\frac{2}{3};\frac{4}{3},\frac{5}{3};-\frac{1}{27 z}\right)}{2 z^{2/3} \Gamma \left(\frac{1}{3}\right)^2}+\frac{\Gamma \left(\frac{1}{3}\right) \, _3F_2\left(\frac{1}{3},\frac{1}{3},\frac{1}{3};\frac{2}{3},\frac{4}{3};-\frac{1}{27 z}\right)}{\sqrt[3]{z} \Gamma \left(\frac{2}{3}\right)^2}\right)-\pi \right), \\
&F_1(z)=-{1\over 2}\log \left({d t(z,0)\over  dz}\right)-\frac{1}{12} \log \left(z^8+\frac{z^7}{27}\right).
\ea\ee
where $z=\kappa^{-3}$. 
In the computation of the Fermionic spectral traces one uses
\be t(z, 2 \pi)= t(-z,0), \quad \partial_t \widehat F_0(z)= \partial_t F_0(-z), \quad \widehat F_1(z)=F_1(-z).\ee
However \eqref{tor}, \eqref{F0ro}  have branch cuts for $z<0$, hence it is more convenient to first do the analytic continuation of  
$\Xi(-\kappa, \hbar)$ and then change $\kappa \to -\kappa$.

A closed form expression for the constant map contribution of this geometry is given in \cite{ghm} and reads
\be
A(\hbar) ={3 A_{\rm c}(\hbar/\pi)- A_{\rm c}(3\hbar/\pi) \over 4},
\ee
where
\be
A_{\rm c}(k)= \frac{2\zeta(3)}{\pi^2 k}\left(1-\frac{k^3}{16}\right)
+\frac{k^2}{\pi^2} \int_0^\infty \frac{x}{\re^{k x}-1}\log(1-\re^{-2x})\rd x.
\ee
A careful evaluation of the integral leads to
 \be A'(2 \pi)=-\frac{{\rm Im}\left(\text{Li}_2\left(\frac{\re^{-\frac{\ri \pi}{6} }}{\sqrt{3}}\right)\right)}{4 \pi ^2}+\frac{2}{27 \sqrt{3}}+\frac{\log (3)}{48 \pi }-\frac{\psi ^{(1)}\left(\frac{1}{3}\right)}{12 \sqrt{3} \pi ^2}, \ee
 where $\psi$ denotes the polygamma function of order 1.

\section{The spectral determinant at the third and fourth order }\label{apo3}
Let us illustrate the cancellation mechanism explained in subsection \ref{qftg} at order 3.
We have
\be\label{full3}\ba  {d^3 \over d \hbar^3}  \Xi(\mu,\hbar)\Big|_{\hbar=2\pi}=& \sum_{n \in \IZ}\re^{J(\mu +2\pi n \ri ,2\pi)}\left((\partial_{\hbar}J(\mu +2\pi n \ri ,2\pi))^3+\partial_{\hbar}^3J(\mu +2\pi n \ri ,2\pi)\right.\\
&+3\partial_{\hbar}^2J(\mu +2\pi n \ri ,2\pi)\partial_{\hbar}J(\mu +2\pi n \ri ,2\pi)  \\
& +{6 }{\widehat  T^{(1)}}\Big( \partial_\hbar \partial_{T} J(\mu +2\pi n \ri ,2\pi)+ \partial_\hbar  J(\mu +2\pi n \ri,2\pi)\partial_T  J(\mu +2\pi n \ri,2\pi)\Big)\Big),\\
\ea \ee
where 
\be \label{J3}\ba \partial_{\hbar}^3J(\mu +2\pi n \ri ,2\pi)\Big|_{\hbar=2 \pi}=&-\frac{1}{20} \ri \left(n-{\widehat T \over 2\pi \ri}\right)^5 \partial_T^5  \widehat F_0+\frac{1}{4 \pi }{\left(n-{\widehat T \over 2\pi \ri}\right)^4 \partial_T^4  \widehat F_0} +\frac{\ri}{4 \pi ^2}{ \left(n-{\widehat T \over 2\pi \ri}\right)^3 \partial_T^3  \widehat F_0}
\\
&-\ri \left(n-{\widehat T \over 2\pi \ri}\right)^3 \partial_T^3 \widehat F_1+\frac{3 }{\pi }\left(n-{\widehat T \over 2\pi \ri}\right)^2 \partial_T^2 \widehat F_1+\frac{3 \ri}{2 \pi ^2}{\left(n-{\widehat T \over 2\pi \ri}\right) \partial_T \widehat F_1}
\\& -6 \ri \left(n-{\widehat T \over 2\pi \ri}\right) \partial_T  \widehat F_2+\frac{6 }{\pi }  \widehat F_2+6 \ri \left(n-{\widehat T \over 2\pi \ri}\right) \partial_T  \widehat F_2^{\rm NS}-\frac{6 }{\pi }  \widehat F_2^{\rm NS}.\ea\ee
The others quantities  in \eqref{full3} have been computed in section \ref{qtf}.
As explained in subsection \ref{qftg}, \eqref{full3} can be written as a part involving only standard free energies, which is modular invariant, plus two extra prices \eqref{NN1} and \eqref{NN2}. In this case the term \eqref{NN1}  reads 
\be \label{3first}\ba
   \re^{N^2  {\widehat F_0} +\widehat F_1-\widehat F_1^{\rm NS}}&\Big(-\frac{6}{\pi}\widehat   F_2^{\rm  NS}\Theta^{(0)}+\ri \left(-18  \widehat  F_2^{\rm NS} \partial_T \widehat F_1-6  \widehat  F_2 \partial_T \widehat F_1^{\rm NS}-18  \widehat  F_2^{\rm NS} \partial_T \widehat F_1^{\rm NS}+6  \partial_T \widehat  F_2^{\rm NS}\right)\Theta^{(1)}+ \\
&-3{\ri } \left( \partial_T^3  {\widehat F_0}\widehat F_2^{\rm NS}+\partial_T^2\widehat F_1 \partial_T \widehat F_1^{\rm NS}+  \partial_T \widehat F_1 (\partial_T \widehat F_1^{\rm NS})^2+ (\partial_T \widehat F_1)^2 \partial_T \widehat F_1^{\rm NS}+ {1\over 3} (\partial_T \widehat F_1^{\rm NS})^3\right)\Theta^{(3)}\\
&+\frac{\ri }{12 } \left(-3   \partial_T^4  {\widehat F_0} \partial_T \widehat F_1^{\rm NS}-12  \partial_T^3  {\widehat F_0} \partial_T \widehat F_1 \partial_T \widehat F_1^{\rm NS}-6  \partial_T^3  {\widehat F_0} ( \partial_T \widehat F_1^{\rm NS})^2\right)\Theta^{(5)}\\
&+\frac{3}{\pi } \partial_T \widehat F_1 \partial_T \widehat F_1^{\rm NS}\Theta^{(2)}-\frac{\ri}{12}  (\partial_T^3  {\widehat F_0})^2 \partial_T \widehat F_1^{\rm NS}\Theta^{(7)}+{1\over 2 \pi }{\partial_T^3  {\widehat F_0} \partial_T \widehat F_1^{\rm NS}}\Theta^{(4)}\Big).
\ea\ee
Similarly  the second term \eqref{NN2}  reads
\be \label{3second} \ba 
  \re^{N^2  {\widehat F_0} +\widehat F_1-\widehat F_1^{\rm NS}} &\Big(-\frac{3\widehat T^{(1)}}{\pi }\left(\partial_T \widehat F_1^{\rm NS}+\partial_T \widehat F_1\right)\Theta^{(0)}-\frac{3 \widehat T^{(1)} }{2 \pi }\partial_T^3  {\widehat F_0}\Theta^{(2)}+\frac{1}{2} \ri \widehat T^{(1)}(\partial_T^3  {\widehat F_0})^2\Theta^{(5)}\\
&+ 6 \ri \widehat T^{(1)} \left(\partial_T^2 \widehat F_1+(\partial_T \widehat F_1)^2+\partial_T^2 \widehat F_1^{\rm NS}-(\partial_T \widehat F_1^{\rm NS})^2\right)\Theta^{(1)}\\
&+ \ri \widehat T^{(1)} \left(\partial_T^4  {\widehat F_0}+2 \partial_T^3  {\widehat F_0} \left(2 \partial_T \widehat F_1+\partial_T \widehat F_1^{\rm NS}\right)\right)\Theta^{(3)}\Big).
\ea\ee
We can carefully compute the symplectic transformation of \eqref{NN2} and \eqref{NN1} by using the modular transformations worked out in the section \ref{mns} together with   \eqref{tt} and 
\be \label{F2tras} \ba \widehat F_2 \to &\widehat F_2+{1\over 2}{  {\mathcal K} \over 2 \pi \ri}\left((\partial_T \widehat F_1)^2+\partial_T^2 \widehat F_1\right)+{1\over 8}{ \left({\mathcal K}\over 2 \pi \ri \right)^2}\left(4\partial_T \widehat F_1\partial_T^3 \widehat F_0+\partial_T^4 \widehat F_0\right)+\left({\mathcal K} \over 2 \pi \ri\right)^3{5 \over 24} (\partial_T^3 \widehat F_0)^2.\ea \ee
It follows that both the terms \eqref{3second} and \eqref{3first}  transform into itself with a shift of ($\pm$) the following factor:
\be \ba  \re^{N^2  {\widehat F_0} +\widehat F_1-\widehat F_1^{\rm NS}}&\Big(-\frac{3 {\mathcal K} }{2 \ri \pi^2}{  \partial_T \widehat F_1^{\rm NS} \left(\partial_T \widehat F_1+\partial_T \widehat F_1^{\rm NS}\right)}\Theta^{(0)} -\frac{3{\mathcal K} }{4 \ri \pi^2 }{ \partial_T^3  {\widehat F_0}  \partial_T \widehat F_1^{\rm NS}}\Theta^{(2)}+\frac{ {\mathcal K} }{4 \pi}  (\partial_T^3  {\widehat F_0})^2  \partial_T \widehat F_1^{\rm NS}\Theta^{(5)}\\
&+{3\over \pi}   {\mathcal K} \partial_T \widehat F_1^{\rm NS} \left(\partial_T^2 \widehat F_1+(\partial_T \widehat F_1)^2+\partial_T^2 \widehat F_1^{\rm NS}-(\partial_T^2 \widehat F_1^{\rm NS})^2\right)\Theta^{(1)}\\
& {1\over 2 \pi}   {\mathcal K} \partial_T \widehat F_1^{\rm NS} \left(\partial_T^4  {\widehat F_0}+2 \partial_T^3  {\widehat F_0} \left(2 \partial_T \widehat F_1+\partial_T \widehat F_1^{\rm NS}\right)\right)\Theta^{(3)}\Big).\ea \ee
Therefore when we add  \eqref{3second} and \eqref{3first} this cancels and we obtain modular invariance also at the third order with the mechanism described below equation \eqref{NN2}. 

Similarly we have
\be\label{full4}\ba  {d^4 \over d \hbar^4}  \Xi(\mu,\hbar)&\Big|_{\hbar=2\pi}=
\sum_{n \in \IZ}\re^{J(\mu +2\pi n \ri ,2\pi)}\left( \partial_\hbar J(\mu+2\pi\ri n, 2\pi)^4+6 \partial_\hbar^2 J(\mu+2\pi\ri n, 2\pi) \partial_\hbar J(\mu+2\pi\ri n, 2\pi)^2+ \right.\\
&3 \partial_\hbar^2 J(\mu+2\pi\ri n, 2\pi)^2+4 \partial_\hbar J(\mu+2\pi\ri n, 2\pi) \partial_\hbar^3 J(\mu+2\pi\ri n, 2\pi)+\partial_\hbar^4 J(\mu+2\pi\ri n, 2\pi)+\\
&6 \widehat T^{(2)} \partial_T J(\mu+2\pi\ri n, 2\pi)+(\widehat T^{(1)})^2 \left(12 \partial_T J(\mu+2\pi\ri n, 2\pi)^2+12 \partial_T^2 J(\mu+2\pi\ri n, 2\pi)\right)+ \\
&\widehat T^{(1)} \left(12 \partial_T J(\mu+2\pi\ri n, 2\pi) \partial_\hbar J(\mu+2\pi\ri n, 2\pi)^2
+12 \partial_T \partial_\hbar^2 J(\mu+2\pi\ri n, 2\pi)
\right) +\\
&12\widehat T^{(1)} \left( \partial_\hbar^2 J(\mu+2\pi\ri n, 2\pi) \partial_T J(\mu+2\pi\ri n, 2\pi)+2 \partial_T \partial_\hbar J(\mu+2\pi\ri n, 2\pi) \partial_\hbar J(\mu+2\pi\ri n, 2\pi)\right).
\ea \ee
By using 
\be \ba&  \partial^4_{\hbar} J(\mu+2\pi n \ri, \hbar)\Big|_{\hbar=2 \pi}= -\frac{18  \widehat F_2}{\pi ^2}+24  \widehat F_3-120  \widehat F_3^{\rm NS} +\left(n-{\widehat T \over 2\pi \ri}\right) \left(\frac{36 \ri \partial_T \widehat F_2}{\pi }-\frac{3 \ri  \partial_T \widehat F_1}{\pi ^3} \right) \\
& +\left(n-{\widehat T \over 2\pi \ri}\right)^2 \left(12 \partial_T^2 \widehat F_2-\frac{9 \partial_T^2 \widehat F_1}{\pi ^2} \right)  +\left(n-{\widehat T \over 2\pi \ri}\right)^3 \left(\frac{6 \ri \partial_T^3 \widehat F_1}{\pi }-\frac{\ri \partial_T^3 \widehat F_0}{2 \pi ^3}\right)\\
&+ \frac{1}{30} \left(n-{\widehat T \over 2\pi \ri}\right)^6\partial_T^6 \widehat F_0+\left(n-{\widehat T \over 2\pi \ri}\right)^5\frac{3 \ri \partial_T^5 \widehat F_0}{10 \pi }+\left(n-{\widehat T \over 2\pi \ri}\right)^4\left(\partial_T^4 \widehat F_1-\frac{3  \partial_T^4 \widehat F_0}{4 \pi ^2}\right),\ea \ee%
we have checked that the cancellation mechanism explained above  holds also at order 4 and we have indeed symplectic invariance of \eqref{full4}.
\bibliographystyle{JHEP}
\bibliography{biblio}
\end{document}